\documentclass[journal]{IEEEtran}  

\usepackage{amsmath}
\usepackage{amssymb}
\usepackage{amsthm}
\usepackage{mathrsfs}
\usepackage{algorithm}
\usepackage{algorithmic}
\usepackage{setspace}
\usepackage{array}
\usepackage{url,times,color,graphicx,epsfig,cite,psfrag,dsfont,booktabs}
\usepackage{float}
\usepackage{lmodern}  
\usepackage{hyperref}  
\usepackage{newtxtext,newtxmath}
\usepackage{lettrine}
\usepackage{tabularx} 
\usepackage{xcolor}
\usepackage{booktabs}    
\usepackage{multirow}    
\usepackage{makecell}    
\usepackage{amsmath}     

\newcommand{\bm}[1]{\mbox{\boldmath{$#1$}}}

\makeatletter

\newcommand{\Rmnum}[1]{\expandafter\@slowromancap\romannumeral #1@}
\makeatother

\renewcommand{\textcolor}[2]{#2}

\begin{document}

\title{
Beamforming and Phase Shift Design for STAR-RIS  Assisted
Secure Sensing and Communication in ISAC Systems
}

\author{Haijun Zhang,~\IEEEmembership{Fellow,~IEEE},
Shuqing Wu, 
Xiaoqi Zhang,~\IEEEmembership{Member,~IEEE},
Zijun Wu,
Xu Ma,
and Yuzheng Ren,~\IEEEmembership{Member,~IEEE}
\thanks{
This work is supported in part by the National Natural Science Foundation of China under Grant 62225103, U22B2003, U2441227, 62561160097, 62541102, and U24A20211, the Beijing
Natural Science Foundation under Grant L253003 and L241008,
the Fundamental Research Funds for the Central Universities under Grant FRF-BD-25-053, and the Xiaomi Fund of Young Scholar.
(\emph{Corresponding author: Haijun Zhang}.)

H. Zhang, S. Wu, X. Zhang, Z. Wu, X. Ma, and Y. Ren 
are with Beijing Key Laboratory of Intelligent and Communication Integration and Hebei Key Laboratory of Space-Air-Ground Intelligent Communication, University of Science and Technology Beijing, Beijing 100083, China
(e-mail: zhanghaijun@ustb.edu.cn; 
wushuqing@xs.ustb.edu.cn;
zhangxiaoqi@ustb.edu.cn;
wuzijun@xs.ustb.edu.cn;
maxu@xs.ustb.edu.cn;
renyuzheng@ustb.edu.cn).
}
}



\maketitle

\begin{abstract}
Integrated sensing and communication (ISAC), as a rapidly advancing technique, 
introduces a fresh approach for achieving secure communication and intelligent sensing for future wireless networks.
An ISAC framework empowered by simultaneously transmitting and reflecting reconfigurable intelligent surfaces (STAR-RIS) is explored in this paper,
where a base station equipped with multiple antennas establishes wireless links to users each with a single antenna during the detection of a point target.
The point target, regarded as an eavesdropper, trying to intercept users' information.
Cramér-Rao bound (CRB) serves as evaluation criterion to assess sensing accuracy of point eavesdropper,
whereas the secrecy rate is employed to quantify the security level of the communication link.
To optimize sensing-communication tradeoff, 
a joint optimization problem is constructed.
To approach the formulated problem, a hybrid Block Coordinate Descent (BCD)-based algorithm is developed, 
which alternately updates the transmission beamforming and STAR-RIS phase shifts,
using successive convex approximation (SCA) technique, penalty dual decomposition (PDD) framework and projected gradient method (PGM).
\end{abstract}


\section{INTRODUCTION}
\IEEEPARstart{T}{he} rapid evolution of wireless networks towards the sixth-generation (6G) \textcolor{blue}{technologies}
has catalyzed researches in integrated sensing and communications (ISAC) technology\cite{ref35}.
By seamlessly coordinating the communicating and sensing functions,
ISAC can improve spectrum efficiency and boost wireless network performance \cite{ref2}, 
making it a key enabler for a range of newly arising applications, such as the Internet of Things (IoT) \cite{ref3}, 
vehicle-to-everything \cite{ref4}, 
and particularly, physical layer security (PLS)\cite{ref5},
which can realize the integration and coordination between the physical and digital worlds simultaneously\cite{ref1}.
Therefore, researchers in academia and industry have been actively investigating the diverse application potential of ISAC.
Many studies have been carried out from various angles, 
such as the waveform designs\cite{ref6}, multi-anntena techniques\cite{ref2}, secure ISAC\cite{ref7} and reconfigurable intelligent surfaces (RIS)-enabled ISAC\cite{ref8, ref9}.

Concurrently, RIS has garnered significant attention for its ability to dynamically control or mainpluate the wireless propagation environment \cite{ref10}.
RIS enhances both sensing and communication capabilities through actively regulating electromagnetic environment.
For communication, RIS coordinates signal's phase adjustment through numerous passive reflecting elements to construct additional transmission paths, enhancing signal strength and link reliability at the receiver while suppressing interference, thus improving spectral efficiency and communication quality.
By tuning phase shifts of individual RIS units, reflection signals are coherently aligned towards an intended direction, leading to an enhancing system achievable rate \cite{ref34}.
For sensing, RIS actively establishes or optimizes line-of-sight (LoS) links with targets, enabling more accurate acquisition of target position, velocity, and other information 
through adjustments to the reflective signal properties in terms of phase and amplitude, especially in complex multipath environments.

Recently, there have been growing interests in the researches of the intergation of RIS and ISAC.
Various investigations have explored RIS-ISAC scenarios under diverse sensing and communication performance criteria.
At first, RIS is employed only for communications, which is near to communication users\cite{ref11, ref12}.
The authors in \cite{ref11} put forward a strategy, 
simultaneously optimizing the constant-modulus waveform and the discrete phase shifts of RIS,
satisfying Cramér-Rao bound (CRB) constraint to reduce interference from multiple users in ISAC system.
In \cite{ref12}, authors achieved the maximization of users' achieveable sum rate respecting to sensing limitations.
To exploit RIS capabilities in sensing-centric applications, 
it is commonly assumed that the RIS is positioned in areas lacking LoS connections between the base station (BS) and users or sensing targets \cite{ref13}–\cite{ref16}.
Authors in \cite{ref32} demonstrated a RIS-aided ISAC scheme, maximizing the system capacity through joint beamforming and phase-shift optimization by proximal policy optimization (PPO) algorithm.

Additionally, the signal-to-interference-plus-noise-ratio (SINR) is frequently adopted, evaluating communication capability.
CRB is normally utilized for evaluating sensing accuracy.
For instance, M. Hua et al. in \cite{ref17} designed an algorithm to design active and passive beamforming to minimize transmission power, 
while fulfilling minimum user SINR threshold and minimize the SINR for the radar.
In particular, within the context of diversified explorations on scenario designs and performance metrics,
the authors in \cite{ref18} specialized in single-user scenarios, 
employing signal-to-noise ratio (SNR) as evaluation criterion for ensuring reliable transmission and accurate detection.
Conversely, authors in \cite{ref19} investigated multi-user scenarios with clutter and proposes an optimization framework to maximize radar SINR under communication rate constraints.
Regarding sensing metrics, parameter estimation has attracted extensive attention.
In \cite{ref20}, the authors formulated and optimized the CRB for angle estimation of point targets, 
along with the response matrix characterization for extended targets.
maximum-likelihood (ML) estimation was employed to obtain the estimated mean squared error (MSE), 
confirming the validity of the CRB-based formulation and optimization.
In \cite{ref21}, RIS was leveraged to suppress the interference of multiple users in target angle estimation under CRB constraints.

Moreover, in ISAC systems, the signals for target detection simultaneously carry confidential information transmitted to communication users\cite{ref9}.
Inevitably, such information may leak to targets, thereby posing potential eavesdropping risks.
Physical layer security technology, leveraging the channel discrepancies between radar and communication, 
has been emerged as an potential anti-eavesdropping solution\cite{ref24}.
For exapmle, the authors of \cite{ref25} proposed three precoding optimization schemes for an ISAC system with one target and one communication user: maximizing secrecy rate, maximizing SINR, and minimizing transmit power.
Q. Liu et al. in \cite{ref26} formulated an achieveable secury rate maximization problem for all the legitimate users in an ISAC network enabled by RIS.
In the proposed scheme, RIS was utilized to enhance secure communication performance without aiding the sensing task,
thereby imposing limitations on the sensing capabilities of ISAC.
In \cite{ref27}, the authors focused on designing secure beamforming strategies of RIS-aided ISAC networks,
with an objective reducing worst-case CRB in estimating the eavesdropper’s 2D direction-of-arrival (DOA).
The above discussions focus on RIS assisting communication or sensing only.
To further verify the capability of RIS in enhancing ISAC,
authors in \cite{ref28} put forward an RIS-empowered secure ISAC scheme, 
where RIS was utilized to enhance secure transmission for authorized users.
The transmission scheme incorporated an additional waveform
to support sensing functionality while minimizing potential data exposure.
In \cite{ref29}, authors designed an IRS-aided unmanned aerial vehicle (UAV) network 
in which a UAV serves a dual-functional BS in support of communication with multiple users and simultaneous object detection, 
and proposed iterative algorithms based on alternating optimization (AO), successive convex approximation (SCA), and manifold optimization (MO) 
to solve a complex secure transmission and energy efficiency problems.

\textcolor{blue}{
While the aforementioned RIS-aided ISAC studies have shown promising performance gains, they are inherently limited by the half-space coverage of conventional RIS. 
As indicated in \cite{ref22}, conventional RIS requires the base station (BS) to be positioned on the same side as the sensing targets or communication users to ensure effective signal coverage. This constraint becomes particularly problematic in ISAC systems where communication users and sensing targets may be geographically separated across different sides of the RIS.
Recently, innovative transceiver designs such as rotatable antennas have been proposed to extend system performance by introducing spatial degrees of freedom (DoFs) through the adjustment of antenna boresights \cite{ref40, ref41}. 
Rotatable antenna enabled systems enhance beam alignment and interference management by optimizing the three-dimensional orientation of each antenna, 
offering significant gains in both communication and sensing applications \cite{ref41}. 
However, akin to conventional RIS, the effectiveness of rotatable antennas is inherently limited to half-space coverage.
The boresight of each antenna primarily benefits users or targets located within its front-facing hemisphere. 
This poses a fundamental challenge for ISAC systems where communication users and sensing targets are likely to reside on opposite sides of a transceiver or infrastructure node. 
To overcome this fundamental limitation and enable full-space coverage, 
the concept of simultaneously transmitting and reflecting reconfigurable intelligent surface (STAR-RIS) has recently emerged \cite{ref23}. Unlike conventional RIS that only reflects signals, STAR-RIS can independently control both transmission and reflection of incident electromagnetic waves. This unique capability makes STAR-RIS exceptionally suitable for ISAC deployments.
}
The ability of STAR-RIS to control signals in both directions makes it highly suitable for ISAC deployments.
In \cite{ref32}, the authors presented a framework that integrated a RIS and a STAR-RIS,
for enabling simultaneous wireless information and power transfer (SWIPT) functionality within multiple-input multiple-output (MIMO)-assisted ISAC frameworks
The target rate was maximized while balancing the communication rate under energy harvesting constraints and phase-shits constraints.
A STAR-RIS-enhanced intelligent transportation system was presented in \cite{ref31}, 
trying to minimize the system delay and loss function of the learning model on the roadside unit (RSU) side.
A terahertz‐band ISAC architecture was develpoed in \cite{ref33},
which synergistically combined stacked intelligent metasurfaces (SIM) for low‐complexity, high‐precision beamforming with STAR‐RIS for full‐space coverage.
The authors developed the meta soft actor-critic (Meta‐SAC) algorithm to jointly optimize power allocation, SIM phase shifts, and STAR‐RIS coefficients.

Nonetheless, most of the existing works focus on communication performance or sensing performance in STAR-RIS-enabled ISAC systems,
neglecting their joint optimization.
Inspired by the prior discussions, this paper aims to investigate a STAR-RIS-empowered ISAC system,
considering sensing and communication tradeoff.
\textcolor{blue}{
The key novelty of employing STAR-RIS in our ISAC framework, as opposed to conventional RIS-based ISAC systems, lies in its ability to provide full-space coverage while enabling simultaneous and independent signal manipulation for both communication and sensing functions. 
In conventional RIS-assisted ISAC, both legitimate users and sensing targets must reside on the same side of the RIS, 
limiting deployment flexibility. 
In contrast, our STAR-RIS-based system naturally partitions the space into two operational zones: a transmission zone for serving communication users and a reflection zone for sensing a point target. 
This unique architecture allows us to jointly design the transmit beamforming at the BS and coefficients at the STAR-RIS to optimize the fundamental trade-off between communication secrecy rate and sensing accuracy，
which has not been thoroughly investigated in prior RIS-ISAC literature.
}

Specifically, aided by an STAR-RIS, the multi-antenna BS transmits information to users with single antenna,
simultaneously sensing a point target, which is probably the eavesdropper.
This paper presents the following key contributions:
\begin{itemize}
    \item A STAR-RIS-empowered ISAC scheme is employed. The overall space is split into two half-zones, one for communicating with users and the other to sense a point target. Specifically, this point target, located at the sensing zone, is the potential eavesdropper attempting to illegally access private information meant for legitimate users.
    \item To achieve performance compromised between sensing and data transmission, a joint problem for transmit beamforming and STAR-RIS phase shifts optimization has been formulated, with an objective of jointly enhancing the secrecy rate while suppressing the CRB of the eavesdropper’s DOA estimation.
    \item The problem involves non-convexity and strong variable coupling, which poses significant challenges for direct optimization. \textcolor{blue}{To this end, a Block Coordinate Descent (BCD) algorithm is designed.} The variables are divided into three parts. For transmit beamforming, a penalty dual decomposition (PDD) framework integrated with SCA is employed. As for STAR-RIS phase shifts, a projected gradient method (PGM) is adopted.
\end{itemize}

The structure of the paper is ordered below.
Section \ref{section2} provides a detailed description of a STAR-RIS-empowered ISAC framework, 
detailing the communication and sensing architecture, signal formulations, and the corresponding channel models.
In Section \ref{section3}, metrics used to assess both communication and sensing are analyzed,
 focusing on the derivation of secrecy rate and the CRB-based angle estimation accuracy.
Section \ref{section4} formulates a joint optimization problem aiming to balance communication security and sensing precision. 
A BCD-based iterative algorithm is developed, where transmission beamforming and STAR-RIS phase shifts can be alternately optimized.
Section \ref{section5} demonstrates simulation-based findings to confirm the efficiency and convergence of proposed framework under various configurations.
Section \ref{section6} summarizes the paper with insights and key findings.

\textbf{Notations}: Throughout the paper, 
italic lowercase letters are used for scalars, bold lowercase letters for vectors, and bold uppercase letters for matrices.
The sets $\mathbb{C}^{N\times M}$ and $\mathbb{R}^{N\times M}$ refer to the spaces of $N\times M$ complex and real-valued matrices. 
For any scalar $a$, $\overline{a}$ and $|a|$ stand for its complex conjugate and absolute value. 
The conjugate transpose, transpose, and conjugate-transpose the of a vector or matrix $\mathbf{A}$ 
is written as $\mathbf{A}^H$, $\mathbf{A}^T$ and $\mathbf{A}^*$. 
The operator $\mathrm{diag}(\mathbf{v})$ forms a diagonal matrix from $\mathbf{v}$. 
$\mathbf{A}\succeq\mathbf{0}$ implies it is positive semidefinite. 
The functions $\mathrm{rank}(\mathbf{A})$ and $\operatorname{tr}(\mathbf{A})$ are used to indicate the rank and trace of $\mathbf{A}$. 
$\mathbb{E}[\cdot]$ and $\Re\{\cdot\}$ extracts expectation operator and real part of a complex argument correspondingly. 
A circularly symmetric complex-valued random variable with mean $\mu$ and power $\sigma^2$ is represented by the distribution $\mathcal{CN}(\mu, \sigma^2)$.

\section{SYSTEM MODEL} \label{section2}
\begin{figure}[!t]
  \centering
  \includegraphics[width=3.5in]{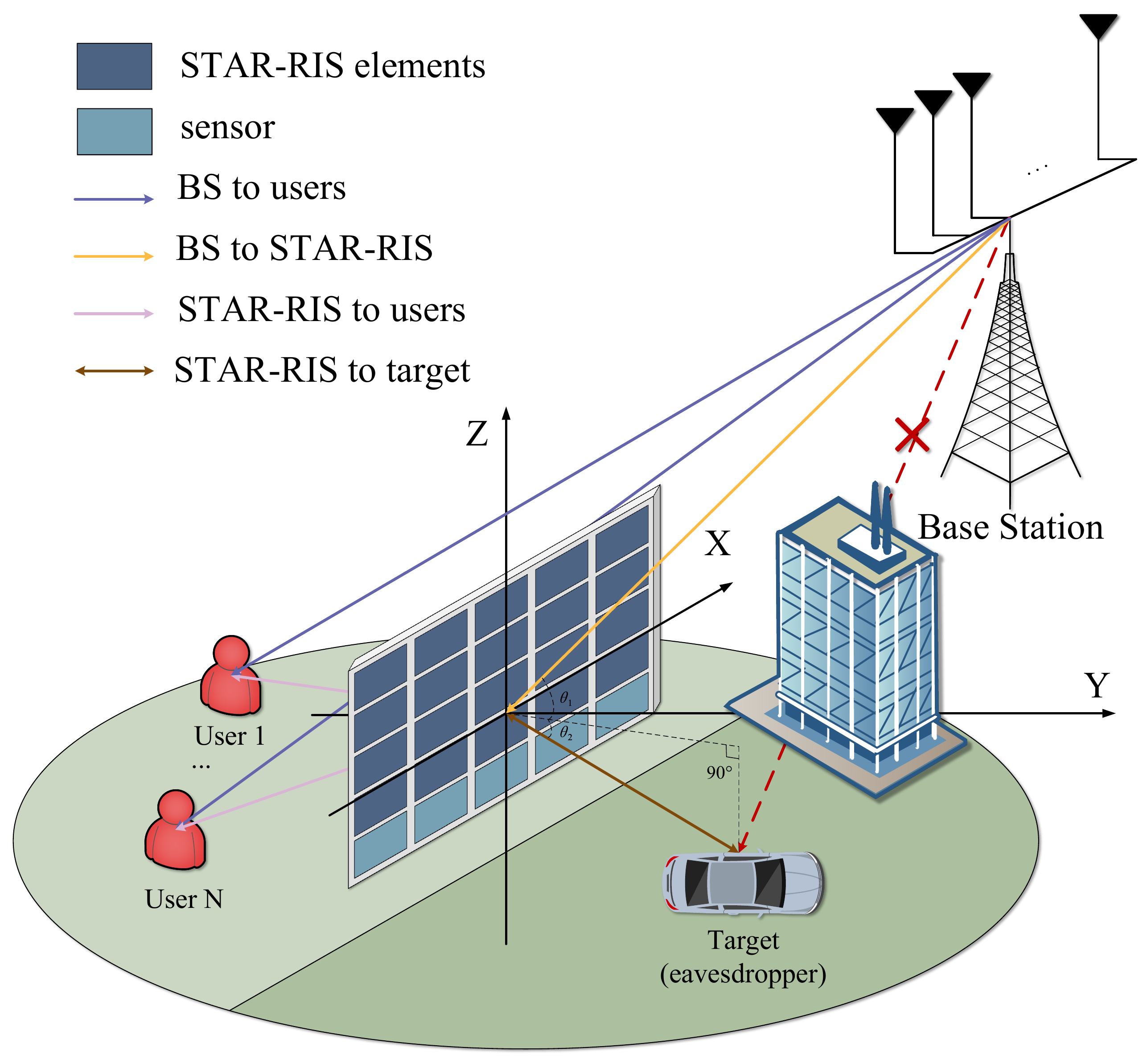}
  \caption{An ISAC system empowered by STAR-RIS}
  \label{fig_1}
\end{figure}

An ISAC system empowered by a STAR-RIS is considered, shown in Fig. \ref{fig_1}.
BS employes a uniform linear array (ULA) with \( M \) antennas,
It is assisted by STAR-RIS deployed in space, 
which consists of a uniform planar array (UPA) \( K \) units, indexed by the set \( \mathcal{K} \).
The STAR-RIS forms two operational zones, a transmission side for communication and a reflection side for sensing.
\textcolor{blue}{
The STAR-RIS comprises two integrated components on the same panel.
A UPA of $K$ reconfigurable elements that simultaneously transmit and reflect incident signals under the energy splitting (ES) protocol.
A ULA of $S$ dedicated sensing antennas, physically adjacent to the reconfigurable UPA, used exclusively for receiving target echoes.
}
Additionally, STAR-RIS is configured with a ULA consisting of $S$ elements functioning as sensing units.
In the communication zone, the system includes \( N \) users, each featuring one single antenna, indexed by \( \mathcal{N} \).
Due to signal degradation along the direct link caused by physical blockages (e.g., buildings), 
the integration of STAR-RIS aims to strengthen overall communication capability.
There is a point target to be detected in the sensing zone, which is the potential eavesdropper for the users.
The direct link from BS to eavesdropper is considered to be impaired by surrounding buildings.
Additionally, an observation period of duration $L$ is defined, 
throughout which the channel coefficients and sensing-related characteristics are treated as static.

\textcolor{blue}{
The BS’s ULA is oriented along the $X$‑axis. 
The STAR-RIS panel lies in the $X$–$Z$ plane with its broadside along the $+Y$ direction. 
The integrated sensing ULA is positioned parallel to the $x$‑axis along one edge of the panel. 
The point target (eavesdropper) is located in the far‑field of the STAR-RIS, with its direction described by azimuth angle $\theta_1$ and elevation angle $\theta_2$ relative to the STAR-RIS broadside. 
Direct links from the BS to the target and from the target to the sensing ULA are assumed blocked by obstacles.
Hence, the target is illuminated solely via the STAR-RIS reflection path.
}

At STAR-RIS, the signal, which is transmitted by the BS, 
is separated into two parts: 
one directed towards the sensing zone, the other towards the communication zone.
Under the ES protocol, every element of STAR-RIS perform signal transmission and reflection simultaneously.
Energy of incident signal is then split between these modes, 
with the splitting ratio determined by the amplitude coefficients \( \beta_{t,k} \) and \( \beta_{r,k} \).
Specifically, \( \beta_{t,k}^2 \) represents the fraction of energy allocated to the transmission mode, while \( \beta_{r,k}^2 \) represents the fraction allocated to the reflection mode,
satisfying the energy conservation constraint in (\ref{eq2}).
Let  \( \mathbf{\Theta}_r \in \mathbb{C}^{K \times K} \) and \( \mathbf{\Theta}_t \in \mathbb{C}^{K \times K} \) denote the transmitting and reflecting coefficient matrices, respectively, 
which is expressed as
\begin{align}
&\mathbf{\Theta}_r = \text{diag}(\beta_{r,1} e^{j\phi_{r,1}}, \dots, \beta_{r,K} e^{j\phi_{r,K}}), \tag {1a} \label{eq1a}\\
&\mathbf{\Theta}_t = \text{diag}(\beta_{t,1} e^{j\phi_{t,1}}, \dots, \beta_{t,K} e^{j\phi_{t,K}}), \tag {1b} \label{eq1b}
\end{align}
where \( \beta_{i,k} \in [0,1]\) and \( \phi_{i,k} \in [0, 2\pi], \forall i \in \left\{t, r\right\}\) represent the amplitude and phase shifts of element \( k \).
\setcounter{equation}{1}

In the model where phase shifts of transmission and reflection are treated independently,
it is postulated that the phases corresponding to transmission and reflection coefficients may be tuned separately. 
However, their respective amplitudes remain subject to the constraint imposed by energy conservation principles
\begin{equation}
\beta_{t,k}^2 + \beta_{r,k}^2 = 1, \quad \forall k \in \mathcal{K}.\label{eq2}
\end{equation}

\textcolor{blue}{
As shown in \cite{ref34},
the low-cost passive and loseless components has been studied to save manufacturing cost of STAR-RIS.
In this case, the transmission and reflection coefficients also need to satisfy the following condition
}
\begin{equation}
\cos(\phi_{t,k} - \phi_{r,k}) = 0, \quad \forall k \in \mathcal{K}. \label{eq3}
\end{equation}
where \( \phi_{t,k} \) and \( \phi_{r,k} \) represent phase shifts of transmission and reflection coefficients, respectively.
\textcolor{blue}{
Equivalently, this implies an orthogonal phase relation $\phi_{t,k} - \phi_{r,k} = \pm \frac{\pi}{2}$,
which is a well-known property of ideal and lossless STAR-RIS elements.
}

\subsection{Communication Signal Model}
At time slot $t$,
the base station transmitted the signal
\textcolor{blue}{
\begin{equation}
  \mathbf{x}(t) = \mathbf{W}\mathbf{c}(t) + \mathbf{s}(t) = \sum_{n = 1}^{N} \mathbf{w}_n c_n(t) + \mathbf{s}(t).
\end{equation}
}

Let \( \mathbf{W} = [\mathbf{w}_1, \dots, \mathbf{w}_N] \in \mathbb{C}^{M \times N} \) to represent the transmit beamforming matrix
corresponding to the data vector \( \mathbf{c}(t) = [c_1(t), \dots, c_N(t)]^T \in \mathbb{C}^{N \times 1} \),
which carries information intended for \( N \) communication users. 
\( \mathbf{s}(t) \in \mathbb{C}^{M \times 1} \) represents the dedicated signal used for sensing purposes, 
characterized by the covariance matrix \( \mathbf{R}_s = \mathbb{E}[\mathbf{s}(t) \mathbf{s}^H(t)] \).
Assume that communication signals are independently drawn from complex Gaussian distributions,
which has zero mean and unit variance, 
satisfying \( \mathbb{E}[\mathbf{c}(t)\mathbf{c}^H(t)] = \mathbf{I}_N \) and \( \mathbb{E}[\mathbf{c}(t)\mathbf{s}^H(t)] = \mathbf{0}_{N \times M} \).

\textcolor{blue}{
Let $L$ denote the total number of observed time samples.
The transmit signal $\mathbf{x}(t)$ characterized by a covariance matrix expressed as
\begin{equation}
  \mathbf{R}_x = \frac{1}{L} \sum_{t=1}^{L}{\mathbf{x}(t)\mathbf{x}^H(t)} \approx
  \mathbb{E}[\mathbf{x}(t)\mathbf{x}^H(t)] = \mathbf{W}\mathbf{W}^H + \mathbf{R}_s.
\end{equation}}

\textcolor{blue}{
Thus, \( \mathbf{R}_x \) is approximated by
\begin{equation}
  \mathbf{R}_x \approx \frac{1}{L} \mathbf{X}\mathbf{X}^H,\label{6}
\end{equation}
where \( \mathbf{X} = [\mathbf{x}(1), \dots, \mathbf{x}(L)] \).
When \( L \) is large enough, the approximation in (\ref{6}) is accurate \cite{ref15}.
}

Let \( \mathbf{g}_n \in \mathbb{C}^{M \times 1} \) denote direct channel between BS and user \( n \).
\( \mathbf{h}_n \in \mathbb{C}^{K \times 1} \) denotes channel from STAR-RIS to user \( n \), 
\( \mathbf{G} \in \mathbb{C}^{K \times M} \) denotes channel from BS to STAR-RIS.
The equivalent channel from BS to user \( n \) is expressed as
\( \mathbf{H}_n = \mathbf{g}_n^H + \mathbf{h}_n^H \mathbf{\Theta}_t \mathbf{G} \).
All the communication channels are modeled using the Rician fading model as follows
\begin{align}
\mathbf{G} &= \frac{\alpha_{\mathbf{G}}}{\sqrt{\epsilon+1}} \left( \sqrt{\epsilon} \mathbf{G}^{\text{LoS}} + \mathbf{G}^{\text{NLoS}} \right), \label{8} \\
\mathbf{h}_n &= \frac{\alpha_{\mathbf{h}_n}}{\sqrt{\epsilon+1}} \left( \sqrt{\epsilon} \mathbf{h}_n^{\text{LoS}} + \mathbf{h}_n^{\text{NLoS}} \right), \label{9} \\
\mathbf{g}_n &= \frac{\alpha_{\mathbf{g}_n}}{\sqrt{\epsilon+1}} \left( \sqrt{\epsilon} \mathbf{g}_n^{\text{LoS}} + \mathbf{g}_n^{\text{NLoS}} \right), \label{10}
\end{align}
where \( \alpha_i, i \in \left\{\mathbf{G}, \mathbf{h}_n, \mathbf{g}_n\right\} \) represents the pathloss,
which is associated with distance, and \( \epsilon \) denotes the Rician factor.
The terms \( \mathbf{G}^{\text{LoS}}, \mathbf{h}_n^{\text{LoS}}, \mathbf{g}_n^{\text{LoS}} \) represent the LoS parts, 
while \( \mathbf{G}^{\text{NLoS}}, \mathbf{h}_n^{\text{NLoS}}, \mathbf{g}_n^{\text{NLoS}} \) represent the non-line-of-sight (NLoS) parts.

The received signal at user \( n \) is formulated as
\begin{equation}
\begin{aligned}
y_{c,n}(t) = & \mathbf{H}_n \mathbf{w}_n \mathbf{c}_n(t) + \mathbf{H}_n \sum_{i = 1, i \neq n}^{N} \mathbf{w}_i \mathbf{c}_i(t) + \mathbf{H}_n \mathbf{s}(t) + n_n(t).
\end{aligned}
\label{11}
\end{equation}

In this expression, the first component represents the desired signal,
the second accounts for interference from other users,
the third reflects the interference by the sensing signal,
and \( n_n(t) \sim \mathcal{CN}(0, \sigma_n^2) \) denotes the additive white Gaussian noise (AWGN) at user \( n \).

\subsection{Sensing Signal Model}
Obviously, eavesdropper receive the signal in the sensing zone
\begin{equation}
  \begin{aligned}
  y_{e}(t) = & \mathbf{h}_e^H \mathbf{\Theta}_r \mathbf{G} \sum_{n = 1}^{N} \mathbf{w}_n \mathbf{c}_n(t) 
  + \mathbf{h}_e^H \mathbf{\Theta}_r \mathbf{G} \mathbf{s}(t)
  + n_e(t),
  \end{aligned}
  \label{12}
\end{equation}
where \( n_e(t) \sim \mathcal{CN}(0, \sigma_e^2) \) denotes the AWGN at the eavesdropper.

The steering vectors of STAR-RIS and sensors are given by
\begin{equation}
\mathbf{a}(\theta_1, \theta_2) = \exp\left(-j \mathbf{R} \mathbf{k}(\theta_1, \theta_2)\right),
\end{equation}
\begin{equation}
\mathbf{b}(\theta_1, \theta_2) = \exp\left(-j \mathbf{\bar{R}} \mathbf{k}(\theta_1, \theta_2)\right),
\end{equation}
where \(\mathbf{R} = [\mathbf{r}_X, \mathbf{r}_Y, \mathbf{r}_Z] \in \mathbb{R}^{K \times 3}\) 
and \(\mathbf{\bar{R}} = [\bar{\mathbf{r}}_X, \bar{\mathbf{r}}_Y, \bar{\mathbf{r}}_Z] \in \mathbb{R}^{S \times 3}\) 
are matrices whose rows represent the Cartesian coordinates of both the STAR-RIS elements and sensors.
The wave vector \( \mathbf{k}(\theta_1, \theta_2) \in \mathbb{R}^{3 \times 1} \) is expressed as
\begin{equation}
\mathbf{k}(\theta_1, \theta_2) = \frac{2\pi}{\lambda_c} 
\begin{bmatrix}
\cos\theta_1 \cos\theta_2 \\
\sin\theta_1 \cos\theta_2 \\
\sin\theta_2
\end{bmatrix},
\end{equation}
where the parameter \( \lambda_c \) specifies the wavelength associated with the carrier wave.
Assuming the ULA sensor array is positioned along the horizontal \( X \)-axis,
and the UPA-configured STAR-RIS is distributed over the \( (X, Z) \) plane. 
Accordingly, the expressions for steering vectors can be reduced to
\begin{equation}
\mathbf{a}(\theta_1, \theta_2) = \exp\left(-j \frac{2\pi}{\lambda_c} \left( \mathbf{r}_X \cos\theta_1 \cos\theta_2 + \mathbf{r}_Z \sin\theta_2 \right)\right),
\end{equation}
\begin{equation}
\mathbf{b}(\theta_1, \theta_2) = \exp\left(-j \frac{2\pi}{\lambda_c} \bar{\mathbf{r}}_X \cos\theta_1 \cos\theta_2 \right).
\end{equation}

\textcolor{blue}{
The BS transmits the signal $\mathbf{X}$,
which is reflected by the STAR-RIS's reconfigurable elements toward the target. 
}
During \( L \) coherent time block,
\textcolor{blue}{
this eavesdropper echo is then received by the co‑located sensing ULA on the STAR-RIS panel, yielding the received signal:
}
\begin{equation}
\mathbf{Y}_s = \beta \mathbf{b}(\theta_1, \theta_2) \mathbf{a}^T(\theta_1, \theta_2) \mathbf{\Theta}_r \mathbf{G} \mathbf{X} + \mathbf{N}_s,
\end{equation}
where \( \beta \in \mathbb{C} \) denotes the complex gain incorporating round-trip path attenuation and reflecting coefficient of eavesdropper;
\( \theta_1 \) and \( \theta_2 \) represent azimuth and elevation angles of eavesdropper relative to STAR-RIS, respectively;
\( \mathbf{a}(\theta_1, \theta_2) \in \mathbb{C}^{K \times 1} \) represents the STAR-RIS steering vector;
\( \mathbf{b}(\theta_1, \theta_2) \in \mathbb{C}^{S \times 1} \) represents the sensor steering vector.
The noise matrix \( \mathbf{N}_s \) comprises entries independently drawn from a circularly symmetric complex Gaussian distribution, which is to be \( \mathcal{CN}(0, \sigma_S^2) \).

\section{PERFORMANCE ANALYSIS} \label{section3}
\subsection{Communication Metric Analysis} 
\textcolor{blue}{
In multi-user communication systems, the SINR is a widely adopted and fundamental metric for quantifying the quality of each communication link. 
It directly captures the impact of both multi-user interference and the dedicated sensing signal, which is essential for evaluating the achievable rate under the ISAC transmission framework. 
Therefore, SINR is adopted to characterize the communication performance for each user.
}Performance of multi-user systems is typically characterized by its sum achievable rate,
which quantifies the maximum rate that is supported for each user under given constraints.
Therefore, this paper concentrates on optimizing sum achievable rate.
The SINR of the desired signal for each user is formulated as
\begin{equation}
\gamma_n = \frac{\left| \mathbf{H}_n \mathbf{w}_n \right|^2}{\sum_{i = 1, i \neq n}^{N} \left| \mathbf{H}_n \mathbf{w}_i \right|^2 + \mathbf{H}_n \mathbf{R}_s \mathbf{H}_n^H + \sigma_n^2},\label{19}
\end{equation}
where \( \mathbf{w}_n \) denotes the beamforming assigned to the user \( n \), 
\( \mathbf{R}_s \) represents the covariance matrix corresponding to the sensing signal, 
while \( \sigma_n^2 \) presents noise power of the user \( n \).
The achievable rate of user $n$ is
\begin{equation}
  \text{R}_n = \log_2(1 + \gamma_n),\label{20}
  \end{equation}
where \( \gamma_n \) is the SINR of the user $n$.

Since the eavesdropper is unable to suppress the interference introduced by the sensing signal,
the corresponding received SINR is given by
\begin{equation}
  \gamma_e = \frac{\left| \mathbf{H}_e \mathbf{w}_n \right|^2}{\sum_{i = 1, i\neq n}^{N} \left| \mathbf{H}_e \mathbf{w}_i \right|^2 + \mathbf{H}_e \mathbf{R}_s \mathbf{H}_e^H + \sigma_e^2},
\end{equation}
where $\mathbf{H}_e=\mathbf{h}_e^H\mathbf{\Theta}_r\mathbf{G}$. 
Then, the eavesdropping rate is
\begin{equation}
  \text{R}_e = \log_2(1 + \gamma_e).\label{22}
\end{equation}

Therefore, secrecy rate, denoted as \( \text{R}_s \), is defined as
\begin{equation}
  \text{R}_s = \text{max}\left\{\sum_{n = 1}^{N}\text{R}_n - \text{R}_e,0\right\}.\label{23}
\end{equation}
\textcolor{blue}{
This work considers an eavesdropper with advanced decoding capability. 
It is assumed that the eavesdropper can jointly process the received signal to decode the superposition of all users' messages from $y_e(t)$ in (\ref{12}). 
Consequently, the SINR $\gamma_e$ characterizes the eavesdropper's effective SINR for decoding the combined information stream, 
and $R_e$ thus represents the corresponding aggregate eavesdropping capacity. 
The system secrecy rate is defined as the maximum achievable rate for reliable communication to all legitimate users while keeping the aggregate information rate at the eavesdropper arbitrarily low. 
Under this model, it is given by the difference between the users' sum-rate and this aggregate eavesdropping capacity \cite{ref37}.}

\subsection{Sensing Metric Analysis}
\textcolor{blue}{
For parameter estimation of a point target, 
the Cramér–Rao Bound provides a theoretical lower bound on the variance of any unbiased estimator. 
It is a widely used benchmark for assessing sensing accuracy, as it is independent of specific estimation algorithms and reflects the fundamental limit imposed by the signal model and noise statistics. 
Thus, minimizing the CRB corresponds to optimizing the best possible sensing performance achievable under the given system configuration.
}For eavesdropper sensing, 
the estimation of \( \theta_1 \) and \( \theta_2 \) according to the observed values \( \mathbf{Z}_s \) over the coherence block 
is performed by MLE.
After obtaining the estimation of \( \hat{\theta}_1 \) and \( \hat{\theta}_2 \), 
the performance is evaluated by MSE
\( \epsilon_1^2 = \mathbb{E}[|\theta_1 - \hat{\theta}_1|^2] \) and \( \epsilon_2^2 = \mathbb{E}[|\theta_2 - \hat{\theta}_2|^2] \).
Obviously, the optimization of the MSE is equivalent to the minimization of the CRB for the parameters \( \theta_1 \) and \( \theta_2 \).
A closed-form expression for CRB is provided, 
serving as a lower bound on MSE.
Therefore, CRB is minimized for estimation of parameters \( \theta_1 \) and \( \theta_2 \).

Firstly, the received signal \( \mathbf{Z}_s \) is vectorized as
\begin{equation}
  \mathbf{z}_s = \text{vec}(\mathbf{Z}_s) = \mathbf{v} + \mathbf{n}_s,\label{24}
\end{equation}
\begin{equation}
  \mathbf{v} = \textcolor{blue}{\beta} \text{vec}(\mathbf{b}(\theta_1, \theta_2) \mathbf{a}^T(\theta_1, \theta_2) \mathbf{\Theta}_r \mathbf{G} \mathbf{X}),\label{25}  
\end{equation}
\begin{equation}
  \mathbf{n}_s = \text{vec}(\mathbf{N}_s). \label{26}
\end{equation}

From equations (\ref{24}) to (\ref{26}), the folllowing can be obtained
\begin{equation}
  \mathbf{y}_s = \textcolor{blue}{\beta} \boldsymbol{\delta}(\theta_1, \theta_2) + \mathbf{n}_s,\label{27}
\end{equation}
\begin{equation}
  \boldsymbol{\delta}(\theta_1, \theta_2) = \text{vec}(\mathbf{b}(\theta_1, \theta_2) \mathbf{a}^T(\theta_1, \theta_2) \mathbf{\Theta}_r \mathbf{G} \mathbf{X}). \label{28}
\end{equation}

A combined vector \( \boldsymbol{\xi} = \begin{bmatrix} \boldsymbol{\theta}^T, \tilde{\boldsymbol{\alpha}}^T \end{bmatrix}^T \) is formulated, 
in which all parameters to be estimated are included,
where \( \boldsymbol{\theta} = \begin{bmatrix} \theta_1, \theta_2 \end{bmatrix}^T \)
and \( \textcolor{blue}{\tilde{\boldsymbol{\beta}}} = \begin{bmatrix} \text{Re}(\textcolor{blue}{\beta}), \text{Im}(\textcolor{blue}{\beta}) \end{bmatrix}^T \).
The vector $\mathbf{z}_s$ is a Gaussian vector,
with the mean of $\alpha$ and the variance of $\sigma_s^2 \mathbf{I}_{LS}$,
which maximum likelihood function is as follows, given parameters \( \boldsymbol{\xi} \)
\begin{equation}
  f(\mathbf{z}_s; \boldsymbol{\xi}) = \frac{1}{(\pi \sigma_s^2)^{LS}} \exp\left( - \frac{\| \mathbf{z}_s - \textcolor{blue}{\beta} \boldsymbol{\delta}(\theta_1, \theta_2) \|^2}{\sigma_s^2} \right).\label{29}
\end{equation}

Thus, MLE of \( \boldsymbol{\xi} \) is obtained 
through solving the next optimization problem
\begin{equation}
  \hat{\boldsymbol{\xi}} = \arg \max_{\boldsymbol{\xi}} f(\mathbf{z}_s; \boldsymbol{\xi}) = 
  \arg \min_{\boldsymbol{\xi}} \| \mathbf{z}_s - \textcolor{blue}{\beta} \boldsymbol{\delta}(\theta_1, \theta_2) \|^2. \label{30} 
\end{equation}

The problem is solved by using numerical methods such as gradient descent or other iterative algorithms,
depending on the complexity of the system model and the steering vectors.
The estimation of \(\textcolor{blue}{\beta} \) for given \( \theta_1 \) and \( \theta_2 \) is obtained as
\begin{equation}
  \textcolor{blue}{\hat{\beta}} = \arg \min_{\textcolor{blue}{\beta}} \| \mathbf{z}_s - \textcolor{blue}{\beta} \boldsymbol{\delta}(\theta_1, \theta_2) \|^2
  = \frac{\boldsymbol{\delta}^H(\theta_1, \theta_2) \mathbf{z}_s}{\|\boldsymbol{\delta}(\theta_1, \theta_2)\|^2}.\label{31}
\end{equation}

Given $\textcolor{blue}{\hat{\beta}}$, what can be obtained is
\begin{equation}
  \left\| \mathbf{z}_s - \textcolor{blue}{\beta} \boldsymbol{\delta}(\theta_1, \theta_2) \right\|^2
  = \left\| \mathbf{z}_s \right\|^2 - \frac{\left| \boldsymbol{\delta}^H(\theta_1, \theta_2) \mathbf{z}_s \right|^2}{\left\| \boldsymbol{\delta}(\theta_1, \theta_2) \right\|^2}.\label{32}
\end{equation}

The MLE of \( \theta_1 \) and \( \theta_2 \) can be obtained as
\begin{equation}
  (\hat{\theta}_1, \hat{\theta}_2)
  = \arg \max_{\theta_1, \theta_2} \frac{\left| \boldsymbol{\delta}^H(\theta_1, \theta_2) \mathbf{z}_s \right|^2}{\|\boldsymbol{\delta}(\theta_1, \theta_2)\|^2}. \label{33}
\end{equation}

It is evident that the observation \(\mathbf{z}_s\) follows a Gaussian distribution.
Given that \( \mathbf{z}_s \sim \mathcal{CN}(\mathbf{v}, \mathbf{R}_n) \),
the \((l, p)\)-th element of Fisher information matrix (FIM) \( \mathbf{J}_\xi \) is
\begin{equation}
  \begin{aligned}
    J_{\boldsymbol{\xi}}(l,p) = & 2 \Re \left\{ \frac{\partial \mathbf{v}^H}{\partial \xi_l} \mathbf{R}_n^{-1} \frac{\partial \mathbf{v}}{\partial \xi_p} \right\} + \operatorname{tr} \left( \mathbf{R}_n^{-1} \frac{\partial \mathbf{R}_n}{\partial \xi_l} \mathbf{R}_n^{-1} \frac{\partial \mathbf{R}_n}{\partial \xi_p} \right) \\
    = & \frac{2}{\sigma_s^2} \Re \left\{ \frac{\partial \mathbf{v}^H}{\partial \xi_l} \frac{\partial \mathbf{v}}{\partial \xi_p} \right\},
  \end{aligned}
  \label{34}
\end{equation}
where $\mathbf{R}_n = \sigma_s^2 \mathbf{I}_{SL}$ represents the convariance matrix for $\mathbf{n}_s$
\( \xi_l \) denotes the \( l \)-th element of \( \boldsymbol{\xi} \).

By defining \( \mathbf{B} = \mathbf{b}(\theta_1, \theta_2) \mathbf{a}^T(\theta_1, \theta_2) \),
there will be
\begin{equation}
\frac{\partial \mathbf{v}}{\partial \boldsymbol{\theta}} = 
\begin{bmatrix}
\textcolor{blue}{\beta} \text{vec}(\dot{\mathbf{B}}_{\theta_1} \mathbf{\Theta}_r \mathbf{G} \mathbf{X}), &
\textcolor{blue}{\beta} \text{vec}(\dot{\mathbf{B}}_{\theta_2} \mathbf{\Theta}_r \mathbf{G} \mathbf{X})
\end{bmatrix},
\tag{34a}
\end{equation}
\begin{equation}
\frac{\partial \mathbf{v}}{\partial \textcolor{blue}{\tilde{\boldsymbol{\beta}}}} = 
\text{vec}(\mathbf{B} \mathbf{\Theta}_r \mathbf{G} \mathbf{X})
\begin{bmatrix}
1 \\ j
\end{bmatrix},
\tag{34b}
\end{equation}
where
\begin{equation}
  \begin{aligned}
  \dot{\mathbf{B}}_{\theta_1} = & \frac{\partial \mathbf{B}}{\partial \theta_1} = \frac{\partial \mathbf{b}}{\partial \theta_1} \mathbf{a}^T + \mathbf{b} \frac{\partial \mathbf{a}^T}{\partial \theta_1}\\
  = & j \frac{2\pi}{\lambda_c} \sin\theta_1 \cos\theta_2 \left( \text{diag}(\bar{\mathbf{r}}_X) \mathbf{b} \mathbf{a}^T + \mathbf{b} \mathbf{a}^T \text{diag}(\mathbf{r}_X) \right),\\
  \end{aligned}
  \label{36}
  \tag{35a}
\end{equation}
\begin{equation}
  \begin{aligned}
    \dot{\mathbf{B}}_{\theta_2} = & \frac{\partial \mathbf{B}}{\partial \theta_2} = \frac{\partial \mathbf{b}}{\partial \theta_2} \mathbf{a}^T + \mathbf{b} \frac{\partial \mathbf{a}^T}{\partial \theta_2} \\
    = & j \frac{2\pi}{\lambda_c} \cos\theta_1 \sin\theta_2 \left( \text{diag}(\bar{\mathbf{r}}_X) \mathbf{b} \mathbf{a}^T + \mathbf{b} \mathbf{a}^T \text{diag}(\mathbf{r}_X) \right) \\
    & - j \frac{2\pi}{\lambda_c} \cos\theta_2 \mathbf{b} \mathbf{a}^T \text{diag}(\mathbf{r}_Z). \label{eq34}
  \end{aligned}
  \tag{35b}
\end{equation}
\addtocounter{equation}{2}

Substituting \( \theta_1 \) and \( \theta_2 \) into \( \mathbf{b}(\theta_1, \theta_2) \) and \( \mathbf{a}(\theta_1, \theta_2) \) for the convenience of the following calculations,
the expression for \( \mathbf{J}_{\boldsymbol{\theta}\boldsymbol{\theta}} \) can be derived as follows
\begin{equation}
  \mathbf{J}_{\theta\theta} = \begin{bmatrix}
    \mathbf{J}_{\theta_1 \theta_1} & \mathbf{J}_{\theta_1 \theta_2} \\
    \mathbf{J}_{\theta_1 \theta_2} & \mathbf{J}_{\theta_2 \theta_2}
    \end{bmatrix},\label{37}
\end{equation}
where
\begin{equation}
  \begin{aligned}
  \mathbf{J}_{\theta_l\theta_p} & =\frac{2}{\sigma_s^2}\Re\left\{\textcolor{blue}{\beta}^*\mathrm{vec}(\dot{\mathbf{B}}_{\theta_l}\mathbf{\Theta}_r\mathbf{G}\mathbf{X})^H\textcolor{blue}{\beta}\mathrm{vec}\left(\dot{\mathbf{B}}_{\theta_p}\mathbf{\Theta}_r\mathbf{G}\mathbf{X}\right)\right\} \\
   & =\frac{2|\textcolor{blue}{\beta}|^2L}{\sigma_s^2}\Re\left\{\operatorname{tr}\left(\dot{\mathbf{B}}_{\theta_p}\mathbf{\Theta}_r\mathbf{G}\mathbf{R}_x\mathbf{G}^H\mathbf{\Theta}_r^H\dot{\mathbf{B}}_{\theta_\ell}^H\right)\right\}.
  \end{aligned}
  \label{37a}
  \tag{37}
\end{equation}

\textcolor{blue}{
The FIM matrix entries include the factor \( L \), 
which originates from substituting \( \mathbf{R}_x = \frac{1}{L} \mathbf{X}\mathbf{X}^H \) into the calculation. 
The step-by-step derivation is as follows:  
From (\ref{25}), the mean vector of the vectorized echo signal $ \mathbf{z}_s $ is $\mathbf{v} = \alpha \cdot \text{vec}\left(\mathbf{B}\Theta_r\mathbf{G}\mathbf{X}\right)$.}

\textcolor{blue}{
To compute the FIM entry \( \mathbf{J}_{\theta_l\theta_p} \), 
the inner product \( \text{vec}^H\left(\dot{\mathbf{B}}_{\theta_l}\Theta_r\mathbf{G}\mathbf{X}\right) \cdot \text{vec}\left(\dot{\mathbf{B}}_{\theta_p}\Theta_r\mathbf{G}\mathbf{X}\right) \) needs to be calculated, 
where \( \dot{\mathbf{B}}_{\theta_l} = \frac{\partial \mathbf{B}}{\partial \theta_l} \).
Given the following matrix vectorization identity:
\begin{equation}
\text{vec}^H(\mathbf{A}\mathbf{C}) \cdot \text{vec}(\mathbf{B}\mathbf{C}) = \text{tr}\left(\mathbf{C}^H \mathbf{A}^H \mathbf{B} \mathbf{C}\right), \label{38} \tag{38}
\end{equation}
let \( \mathbf{A} = \dot{\mathbf{B}}_{\theta_l}\Theta_r\mathbf{G} \), \( \mathbf{B} = \dot{\mathbf{B}}_{\theta_p}\Theta_r\mathbf{G} \), and \( \mathbf{C} = \mathbf{X} \). 
Substituting these variables into the above, it is obtained that  
\begin{equation}
  \begin{aligned}
    & \text{vec}^H\left(\dot{\mathbf{B}}_{\theta_l}\Theta_r\mathbf{G}\mathbf{X}\right) \cdot \text{vec}\left(\dot{\mathbf{B}}_{\theta_p}\Theta_r\mathbf{G}\mathbf{X}\right) \\
& = \text{tr}\left(\mathbf{X}\mathbf{X}^H \cdot \mathbf{G}^H\Theta_r^H\dot{\mathbf{B}}_{\theta_l}^H\dot{\mathbf{B}}_{\theta_p}\Theta_r\mathbf{G}\right)    
  \end{aligned}
  \tag{38a}
\end{equation}}

\textcolor{blue}{
Substituting \( \mathbf{R}_x = \frac{1}{L} \mathbf{X}\mathbf{X}^H \) into the above gives
\begin{equation}
  \begin{aligned}
    & \text{tr}\left(\mathbf{X}\mathbf{X}^H \mathbf{G}^H\Theta_r^H\dot{\mathbf{B}}_{\theta_l}^H\dot{\mathbf{B}}_{\theta_p}\Theta_r\mathbf{G}\right) \\
    = & L\cdot \text{tr}\left(\mathbf{R}_x \mathbf{G}^H\Theta_r^H\dot{\mathbf{B}}_{\theta_l}^H\dot{\mathbf{B}}_{\theta_p}\Theta_r\mathbf{G}\right).\\
  \end{aligned}
  \tag{39}
\end{equation}}

\textcolor{blue}{
Rearranging the variable \( \mathbf{F} \) gives
\begin{equation}
  \begin{aligned}
    \mathbf{G}^H\Theta_r^H \dot{\mathbf{B}}_{\theta_l}^H\dot{\mathbf{B}}_{\theta_p} \Theta_r\mathbf{G} 
    = \mathbf{G}^H\Theta_r^H \dot{\mathbf{B}}_{\theta_l}^H\dot{\mathbf{B}}_{\theta_p} \frac{\mathbf{F}}{\mathbf{G}\mathbf{R}_x\mathbf{G}^H} \Theta_r\mathbf{G}.
    \end{aligned}
    \tag{40}
\end{equation}}

\textcolor{blue}{
Substituting this into the trace term in (\ref{39}) and 
simplifying via cyclic permutation of trace \( \text{tr}(\mathbf{AB}) = \text{tr}(\mathbf{BA}) \), it yields:
\begin{equation}
  \begin{aligned}
    \text{tr}\left(\mathbf{R}_x  \mathbf{G}^H\Theta_r^H\dot{\mathbf{B}}_{\theta_l}^H\dot{\mathbf{B}}_{\theta_p}\Theta_r\mathbf{G}\right) = \text{tr}\left(\dot{\mathbf{B}}_{\theta_p}\mathbf{F}\dot{\mathbf{B}}_{\theta_l}^H\right).\\
  \end{aligned}
  \label{41}
  \tag{41}
\end{equation}}
  
\textcolor{blue}{
According to (\ref{37a}),
the FIM entry \( \mathbf{J}_{\theta_l\theta_p} \) is:  
\begin{equation}
  \mathbf{J}_{\theta_l\theta_p} 
  = \frac{2|\beta|^2}{\sigma_s^2} \mathfrak{R}\left\{ \text{vec}^H\left(\dot{\mathbf{B}}_{\theta_l}\Theta_r\mathbf{G}\mathbf{X}\right) \text{vec}\left(\dot{\mathbf{B}}_{\theta_p}\Theta_r\mathbf{G}\mathbf{X}\right) \right\}. \tag{42}
\end{equation}}

\textcolor{blue}{
Substituting (\ref{40}) and (\ref{41}) into the above expression explicitly introduces the factor \( L \):  
\begin{equation}
  \mathbf{J}_{\theta_l\theta_p} = \frac{2|\beta|^2 L}{\sigma_s^2} \mathfrak{R}\left\{ \text{tr}\left(\dot{\mathbf{B}}_{\theta_p}\mathbf{F}\dot{\mathbf{B}}_{\theta_l}^H\right) \right\}. \tag{43}
\end{equation}
}

Therefore, the matrix \(\mathbf{J}_{\boldsymbol{\theta}\widetilde{\boldsymbol{\alpha}}}\) can be expressed as
\begin{equation}
  \begin{aligned}
    \mathbf{J}_{\boldsymbol{\theta}\widetilde{\boldsymbol{\textcolor{blue}{\beta}}}} & =\frac{2}{\sigma_s^2}\Re\left(
      \begin{bmatrix}
        \textcolor{blue}{\beta}^*\operatorname{vec}(\dot{\mathbf{B}}_{\theta_1}\boldsymbol{\Theta}_r\mathbf{G}\mathbf{X})^H \\
        \textcolor{blue}{\beta}^*\operatorname{vec}(\dot{\mathbf{B}}_{\theta_2}\boldsymbol{\Theta}_r\mathbf{G}\mathbf{X})^H
      \end{bmatrix}\operatorname{vec}(\mathbf{B}\boldsymbol{\Theta}_r\mathbf{G}\mathbf{X})[1,j]\right) \\
      & =\frac{2L}{\sigma_s^2}\Re\left(
      \begin{bmatrix}
        \textcolor{blue}{\beta}^*\operatorname{tr}(\mathbf{B}\boldsymbol{\Theta}_r\mathbf{G}\mathbf{R}_x\mathbf{G}^H\boldsymbol{\Theta}_r^H\dot{\mathbf{B}}_{\theta_1}^H) \\
        \textcolor{blue}{\beta}^*\operatorname{tr}(\mathbf{B}\boldsymbol{\Theta}_r\mathbf{G}\mathbf{R}_x\mathbf{G}^H\boldsymbol{\Theta}_r^H\dot{\mathbf{B}}_{\theta_2}^H)
      \end{bmatrix}[1,j]\right),
  \end{aligned}
  \label{38a}
  \tag{38a}
\end{equation}
\begin{equation}
  \begin{aligned}
    \mathbf{J}_{\tilde{\boldsymbol{\textcolor{blue}{\beta}}}\tilde{\boldsymbol{\textcolor{blue}{\beta}}}} = & \frac{2}{\sigma_s^2} \Re\left( \left( \text{vec}(\mathbf{B}\mathbf{\Theta}_r \mathbf{G} \mathbf{X})[1,j] \right)^H \text{vec}(\mathbf{B}\mathbf{\Theta}_r \mathbf{G} \mathbf{X})[1,j] \right) \\
    = & \frac{2}{\sigma_s^2} \Re\left(
      \begin{bmatrix}
        1 \\ j
      \end{bmatrix}^H 
      \begin{bmatrix}
        1 \\ j
      \end{bmatrix} 
    \text{vec}^H(\mathbf{B}\mathbf{\Theta}_r \mathbf{G} \mathbf{X}) \text{vec}(\mathbf{B}\mathbf{\Theta}_r \mathbf{G} \mathbf{X}) \right) \\
    = & \frac{2L}{\sigma_s^2} \mathbf{I}_2 \operatorname{tr}\left( \mathbf{B}\mathbf{\Theta}_r \mathbf{G} \mathbf{R}_x \mathbf{G}^H \mathbf{\Theta}_r^H \mathbf{B}^H \right).
  \end{aligned}
  \label{38b}
  \tag{38b}
\end{equation}
\addtocounter{equation}{2}

Therefore, the FIM to estimate the vector \( \boldsymbol{\xi} \) is given by
\begin{equation}
  \mathbf{J}_{\boldsymbol{\xi}} = 
  \begin{bmatrix}
    \mathbf{J}_{\boldsymbol{\theta}\boldsymbol{\theta}} & \mathbf{J}_{\boldsymbol{\theta}\tilde{\boldsymbol{\textcolor{blue}{\beta}}}} \\
    \mathbf{J}_{\boldsymbol{\theta}\tilde{\boldsymbol{\textcolor{blue}{\beta}}}}^T & \mathbf{J}_{\tilde{\boldsymbol{\textcolor{blue}{\beta}}}\tilde{\boldsymbol{\textcolor{blue}{\beta}}}}
  \end{bmatrix}.
  \label{39}
\end{equation}

Therefore, to estimate \( \boldsymbol{\theta} \),
the CRB martix is expressed as
\begin{equation}
  \text{CRB}(\boldsymbol{\theta}) = \left[ \mathbf{J}_{\boldsymbol{\theta}\boldsymbol{\theta}} - \mathbf{J}_{\boldsymbol{\theta}\tilde{\boldsymbol{\alpha}}} \mathbf{J}_{\tilde{\boldsymbol{\alpha}}\tilde{\boldsymbol{\alpha}}}^{-1} \mathbf{J}_{\boldsymbol{\theta}\tilde{\boldsymbol{\alpha}}}^T \right]^{-1}.
  \label{40}
\end{equation}

It is apparent that the CRB (\( \boldsymbol{\theta} \)) 
is highly influenced by the eavesdropper DOAs, denoted by \( \boldsymbol{\theta} \).
Therefore, designing both the beamforming and the phase shifts,
optimizing \( \boldsymbol{\theta} \) is crucial.
In practice, eavesdropper DOAs remain nearly unchanged between successive coherent intervals.
Thus, the DOAs estimated or predicted from the previous $\mathbf{\theta}$ coherent time block are feasible.
Based on the above knowledge,
it is assumed that the DOAs of $ \mathbf{\theta}$ in the following optimization problem are fixed.

\section{SECRECY RATE AND SECRCECY RATE TRADEOFF} \label{section4}
The optimization problem of maximizing secrecy rate while minimizing CRB matrix of the eavesdropper DOAs is elaborated on.
Based on the proposed framework, the problem can be solved by a BCD algorithm,
trying to achieve the system performance balance.
\subsection{Problem Formulation}
\textcolor{blue}{
In a STAR-RIS-assisted ISAC system where the sensing target also acts as a potential eavesdropper, there exists an inherent trade-off between communication security and sensing accuracy. 
On one hand, improving the secrecy rate requires steering signals away from the eavesdropper to minimize information leakage.
On the other hand, enhancing sensing precision demands focusing signal energy toward the target to obtain a lower CRB for DOA estimation. 
}To achieve tradeoff between CRB matrix and secrecy rate,
the aim is to enhance the secrecy rate ,
while simultaneously reducing CRB matrix of eavesdropper DOAs. 
Obviously, \( \text{tr}(\mathbf{U}^{-1}) \) exhibits a matrix-decreasing property over the set of positive semidefinite matrices.
Since \( \mathbf{J}_{\boldsymbol{\theta}\boldsymbol{\theta}} - \mathbf{J}_{\boldsymbol{\theta}\tilde{\boldsymbol{\alpha}}} \mathbf{J}_{\tilde{\boldsymbol{\alpha}}\tilde{\boldsymbol{\alpha}}}^{-1} \mathbf{J}_{\boldsymbol{\theta}\tilde{\boldsymbol{\alpha}}}^T \)
is a positive semidefinite matrix,
minimizing \( \text{tr}(\mathbf{J}_{\boldsymbol{\theta}\boldsymbol{\theta}} - \mathbf{J}_{\boldsymbol{\theta}\tilde{\boldsymbol{\alpha}}} \mathbf{J}_{\tilde{\boldsymbol{\alpha}}\tilde{\boldsymbol{\alpha}}}^{-1} \mathbf{J}_{\boldsymbol{\theta}\tilde{\boldsymbol{\alpha}}}^T) \)
is the same as minimizing \( \text{tr}(\mathbf{U}^{-1}) \),
defining an auxiliary martix \( \mathbf{U} \in \mathbb{C}^{2 \times 2} \),
subject to the constraint \(\quad \mathbf{J}_{\boldsymbol{\theta}\boldsymbol{\theta}} - \mathbf{J}_{\boldsymbol{\theta}\tilde{\boldsymbol{\alpha}}} \mathbf{J}_{\tilde{\boldsymbol{\alpha}}\tilde{\boldsymbol{\alpha}}}^{-1} \mathbf{J}_{\boldsymbol{\theta}\tilde{\boldsymbol{\alpha}}}^T \succeq \mathbf{U}\).
Thus, the trace of the CRB matrix minimization is reformulated as the following problem

\begin{align}
  \mathcal{P}: & \max_{\mathbf{U}, \mathbf{W}, \mathbf{R}_s, \boldsymbol{\theta}_t, \boldsymbol{\theta}_r}
  \frac{\mu}{q_s}(\sum_{n\in\mathcal{N}}\text{R}_n-\text{R}_e) + \frac{1-\mu}{q_c}\left[-\text{tr}(\mathbf{U}^{-1})\right],  \\
  \text{s.t.} \quad & \begin{bmatrix}
    \mathbf{J}_{\boldsymbol{\theta}\boldsymbol{\theta}} - \mathbf{U} & \mathbf{J}_{\boldsymbol{\theta}\tilde{\boldsymbol{\alpha}}} \\
    \mathbf{J}_{\boldsymbol{\theta}\tilde{\boldsymbol{\alpha}}}^T & \mathbf{J}_{\tilde{\boldsymbol{\alpha}}\tilde{\boldsymbol{\alpha}}}
  \end{bmatrix} \succeq 0, \tag{41a} \label{41a} \\
  & \text{tr}(\mathbf{W}\mathbf{W}^H + \mathbf{R}_s) \leq P, \tag{41b} \label{41b} \\
  & \mathbf{R}_s \succeq 0, \mathbf{U} \succeq 0, \tag{41c} \label{41c} \\
  & \beta_{t,k}^2 + \beta_{r,k}^2 = 1, \textcolor{blue}{0 \le \beta_{t,k}, \beta_{r,k} \le 1}, \forall k, \tag{41d} \label{41d}
\end{align}
where \(\boldsymbol{\theta}_i = \begin{bmatrix} 
  \beta_{i,1} e^{j\phi_{i,1}}, \beta_{i,2} e^{j\phi_{i,2}}, \dots, \beta_{i,K} e^{j\phi_{i,K}} 
  \end{bmatrix}^T, \forall i \in \{t,r\}\)
represents a vector consisting of diagonal elements of \( \boldsymbol{\Theta}_t \) and \( \boldsymbol{\Theta}_r \).
\(P \geq 0\) in (\ref{41b}) indicates BS's total power budget.
The constraint (\ref{41c}) ensures that the matrices \(\mathbf{R}_s\) and \(\mathbf{U}\) are positive semi-definite.
And the constraint (\ref{41d}) represents the amplitude relationship of STAR-RIS coefficients.

\subsection{Beamforming and Phase Shifts Optimization}
A hybrid BCD algorithm is proposed to tackle \(\mathcal{P}\) in this section,
with its overall structure described as follows.
The BCD algorithm updates each variable block in an iterative manner,
while fixing the other blocks, which results in the next subproblems.
\begin{algorithm}[H]
  \caption{BCD Algorithm for the Tradeoff Optimization between Secrecy Rate and CRB}\label{alg:alg1}
  \begin{algorithmic}[1]
    \STATE \textbf{Input}: $\bm{\theta}_r^{(n)}$, $\bm{\theta}_t^{(n)}$. Set the initial index $n = 1$ and convergence tolerance $\epsilon = 1e-4$.
    \REPEAT
      \STATE {Solve the transmit beamforming problem (\ref{45}) with the other vairables fixed, and obtain $\mathbf{W}$ and $\mathbf{R}_s$.}
      \STATE {Solve the transmit beamforming problem (\ref{58}) with he other vairables fixed, and obtain $\mathbf{U}$ and $\mathbf{F}$.}
      \STATE {Solve the STAR-RIS phase shifts problem (\ref{61}) with the other vairables fixed, and obtain $\mathbf{\bm{\theta}}_r$ and $\mathbf{\bm{\theta}}_t$.}
      \UNTIL {the difference between the $n$-th and ($n - 1$)-th objective value drops beneath the given threshold $\epsilon$}
  \end{algorithmic}
\end{algorithm}

\subsubsection{\textbf{Optimization of transmission beamforming}}
With $\mathbf{\theta}_t$ and $\mathbf{\theta}_r$ fixed,
the transmission beamforming optimizating problem is expressed as
\begin{align}
  \mathcal{P}1: & \max_{\mathbf{U}, \mathbf{W}, \mathbf{R}_s}
  \frac{\mu}{q_s}(\sum_{n\in\mathcal{N}}\text{R}_n-\text{R}_e) + \frac{1-\mu}{q_c}\left[-\text{tr}(\mathbf{U}^{-1})\right], \label{42} \\
  \text{s.t.} \quad & \begin{bmatrix}
    \mathbf{J}_{\boldsymbol{\theta}\boldsymbol{\theta}} - \mathbf{U} & \mathbf{J}_{\boldsymbol{\theta}\tilde{\boldsymbol{\alpha}}} \\
    \mathbf{J}_{\boldsymbol{\theta}\tilde{\boldsymbol{\alpha}}}^T & \mathbf{J}_{\tilde{\boldsymbol{\alpha}}\tilde{\boldsymbol{\alpha}}}
  \end{bmatrix} \succeq 0, \tag{42a} \label{41a} \\
  & \text{tr}(\mathbf{R}_s + \mathbf{W}\mathbf{W}^H) \leq P, \tag{42b} \label{42b} \\
  & \mathbf{R}_s \succeq 0, \tag{42c} \label{42c} \\
  & \mathbf{U} \succeq 0. \tag{42d} \label{42d}
\end{align}

By defining an auxiliary variable \(\mathbf{F} = \mathbf{\Theta}_r \mathbf{G} \mathbf{R}_x \mathbf{G}^H \mathbf{\Theta}_r^H \),
and defining $ \bm{\chi} = \left\{\mathbf{U}, \mathbf{W}, \mathbf{R}_s, \mathbf{F}\right\}$,
the subproblem \(\mathcal{P}1\) can be reformulated as
\begin{align}
  \mathcal{P}1: & \max_{\bm{\chi}}
  \frac{\mu}{q_s}(\sum_{n\in\mathcal{N}}\text{R}_n-\text{R}_e) + \frac{1-\mu}{q_c}\left[-\text{tr}(\mathbf{U}^{-1})\right], \label{43} \\
  \text{s.t.} \quad & \mathbf{F} = \mathbf{\Theta}_r \mathbf{G} \mathbf{R}_x \mathbf{G}^H \mathbf{\Theta}_r^H, \tag{43a} \label{43a}\\
  & \begin{bmatrix}
    \mathbf{J}_{\boldsymbol{\theta}\boldsymbol{\theta}}(\mathbf{F}) - \mathbf{U} & \mathbf{J}_{\boldsymbol{\theta}\tilde{\boldsymbol{\alpha}}}(\mathbf{F}) \\
    \mathbf{J}_{\boldsymbol{\theta}\tilde{\boldsymbol{\alpha}}}^T(\mathbf{F}) & \mathbf{J}_{\tilde{\boldsymbol{\alpha}}\tilde{\boldsymbol{\alpha}}}(\mathbf{F})
  \end{bmatrix} \succeq 0, \tag{43b} \label{43b} \\
  & \text{(\ref{42b})-(\ref{42d})}. \tag{43c} \label{43c} 
\end{align}

And the entries of the constraint (\ref{43b}) are as follows
\begin{equation}
\mathbf{J}_{\boldsymbol{\theta}\boldsymbol{\theta}}(\mathbf{F}) = \frac{2|\textcolor{blue}{\beta}|^2 L}{\sigma_s^2} 
\operatorname{Re}\left(
\begin{bmatrix}
\operatorname{tr}(\dot{\mathbf{B}}_{\theta_1} \mathbf{F} \dot{\mathbf{B}}_{\theta_1}^H) & \operatorname{tr}(\dot{\mathbf{B}}_{\theta_1} \mathbf{F} \dot{\mathbf{B}}_{\theta_2}^H) \\
\operatorname{tr}(\dot{\mathbf{B}}_{\theta_1} \mathbf{F} \dot{\mathbf{B}}_{\theta_2}^H) & \operatorname{tr}(\dot{\mathbf{B}}_{\theta_2} \mathbf{F} \dot{\mathbf{B}}_{\theta_2}^H)
\end{bmatrix}
\right),\tag{44a}
\end{equation}
\begin{equation}
\mathbf{J}_{\boldsymbol{\theta}\tilde{\boldsymbol{\textcolor{blue}{\beta}}}}(\mathbf{F}) = \frac{2L}{\sigma_s^2} 
\operatorname{Re}\left(
\begin{bmatrix}
\textcolor{blue}{\beta}^* \operatorname{tr}(\mathbf{B} \mathbf{F} \dot{\mathbf{B}}_{\theta_1}^H) \\
\textcolor{blue}{\beta}^* \operatorname{tr}(\mathbf{B} \mathbf{F} \dot{\mathbf{B}}_{\theta_2}^H)
\end{bmatrix}
[1, j]
\right), \tag{44b}  
\end{equation}
\addtocounter{equation}{1}
\begin{equation}
  \mathbf{J}_{\tilde{\boldsymbol{\textcolor{blue}{\beta}}}\tilde{\boldsymbol{\textcolor{blue}{\beta}}}}(\mathbf{F}) = \frac{2L}{\sigma_s^2} \mathbf{I}_2 \operatorname{tr}(\mathbf{B} \mathbf{F} \mathbf{B}^H). \tag{44c}
\end{equation}

To solve the problem (\ref{43}), 
the variables $ \bm{\chi}$ to be optimized is divided into two blocks,
$\left\{\mathbf{W},\mathbf{R}_s\right\}$ and $\left\{\mathbf{U},\mathbf{F}\right\}$.

With fixed $\left\{\mathbf{U}, \mathbf{F}\right\}$,
the problem $\mathcal{P}1.1$ is expressed as
\begin{align}
  \mathcal{P}1.1: & \max_{\mathbf{W}, \mathbf{R}_s}
  \frac{\mu}{q_s}(\sum_{n\in\mathcal{N}}\text{R}_n-\text{R}_e), \label{45}\\
  \text{s.t.} \quad & \text{tr}(\mathbf{R}_s + \mathbf{W}\mathbf{W}^H) \leq P, \tag{45a} \\
  & \mathbf{R}_s \succeq 0. \tag{45b} 
\end{align}

Obviously, \(\mathcal{P}1.1\) exhibits a non-convex structure.
By defining $\widetilde{\mathbf{H}}_n = \mathbf{H}_n^H\mathbf{H}_n$,
$\text{R}_n$ can be expressed as $\text{R}_n = N_1 -N_2$,
where
\begin{equation}
  N_1 = \text{log}_2\left\{\sigma_n^2 +\text{tr}\left[\widetilde{\mathbf{H}}_n(\mathbf{W}\mathbf{W}^H+\mathbf{R}_s)\right]\right\},
  \tag{46a} 
\end{equation}
\begin{equation}
  N_2 = \text{log}_2\left\{\sigma_n^2 +\text{tr}\left[\widetilde{\mathbf{H}}_n(\mathbf{W}\mathbf{W}^H+\mathbf{R}_s-\mathbf{w}_n\mathbf{w}_n^H)\right]\right\}.
  \tag{46b}
\end{equation}
\addtocounter{equation}{1}

Similarly, by defining $\widetilde{\mathbf{h}}_e=\mathbf{G}^H\mathbf{\Theta}_r^H\mathbf{h}_e\mathbf{h}_e^H\mathbf{\Theta}_r\mathbf{G}$,
$R_e$ can be expressed as $R_e = n_1-n_2$,
where
\begin{equation}
  n_1 = \text{log}_2\left\{\sigma_e^2 +\text{tr}\left[\widetilde{\mathbf{h}}_e(\mathbf{W}\mathbf{W}^H+\mathbf{R}_s)\right]\right\},
  \tag{47a} 
\end{equation}
\begin{equation}
  n_2 = \text{log}_2\left\{\sigma_e^2 +\text{tr}\left[\widetilde{\mathbf{h}}_e(\mathbf{W}\mathbf{W}^H+\mathbf{R}_s-\mathbf{w}_n\mathbf{w}_n^H)\right]\right\}.
  \tag{47b}
\end{equation}

The following demonstrates that global optimal solution of $\mathcal{P}1.1$ 
is achieved through semidefinite relaxation (SDR) method.
Assume $ \mathbf{W} = [\mathbf{w}_1, \dots, \mathbf{w}_N]^T $.
there exist matrices $ \mathbf{R}_x $ and $ \mathbf{R}_s $ such that $ \mathbf{R}_x = \mathbf{W}\mathbf{W}^H + \mathbf{R}_s $ if and only if $ \mathbf{R}_x \succeq \mathbf{W}\mathbf{W}^H = \sum_{n \in N} \mathbf{w}_n \mathbf{w}_n^H $. 
Define the auxiliary variables $ \mathbf{W}_n = \mathbf{w}_n \mathbf{w}_n^H, \forall n$,
satisfying $ \mathbf{W}_n \succeq 0 $ and $ \text{rank}(\mathbf{W}_n) = 1 $. 

Moreover, with the above expression,
$\text{R}_n =N_1-N_2-n_1+n_2$ can be expressed as
\begin{equation}
  N_1 = \text{log}_2\left\{\sigma_n^2 +\text{tr}\left(\widetilde{\mathbf{H}}_n\mathbf{R}_x\right)\right\},
  \tag{48a}
\end{equation}
\begin{equation}
  N_2 = \text{log}_2\left\{\sigma_n^2 +\text{tr}\left[\widetilde{\mathbf{H}}_n(\mathbf{R}_x-\mathbf{W}_n)\right]\right\},
  \tag{48b} 
\end{equation}
\begin{equation}
  n_1 = \text{log}_2\left\{\sigma_e^2 +\text{tr}(\widetilde{\mathbf{h}}_e\mathbf{R}_x)\right\},
  \tag{48c} 
\end{equation}
\begin{equation}
  n_2 = \text{log}_2\left\{\sigma_e^2 +\text{tr}\left[\widetilde{\mathbf{h}}_e(\mathbf{R}_x-\mathbf{W}_n)\right]\right\}.
  \tag{48d}
\end{equation}
\addtocounter{equation}{2}
The SDR formulation of $ \mathcal{P}1.1 $ can then be expressed as
\begin{align}
  \max_{\mathcal{} \mathbf{W}, \mathbf{R}_x, {\left\{\mathbf{W}_n\right\}}_{n \in \mathcal{N}}} 
  & \frac{\mu}{q_s}\sum_{n\in\mathcal{N}}(N_1 - N_2 - n_1 + n_2), \label{49}\\
  \text{s.t.} \quad & \text{tr}(\mathbf{R}_x) \leq P, \tag{49a}\\
  & \mathbf{R}_x \succeq \sum_{n \in N} \mathbf{W}_n, \mathbf{W}_n \succeq 0, \forall n, \tag{49b}\\
  & \mathbf{R}_s \succeq 0. \tag{49c} 
\end{align}

With the given $\overline{\mathbf{R}}_x$ and $\overline{\mathbf{W}}_n$,
the upper bound is obtained from performing first-order Taylor approximation of the concave function.
Thus, the upper bound of $N_2$ is given by
\begin{equation}
  N_2(\mathbf{W}_n,\mathbf{R}_x) \leq \widetilde{N}_2(\mathbf{W}_n,\mathbf{R}_x).
\end{equation}

The upper bound $\widetilde{N}_2(\mathbf{W}_n,\mathbf{R}_x)$ is given by
\begin{align}
  \widetilde{N}_2(\mathbf{W}_n,\mathbf{R}_x) 
 = & N_2(\overline{\mathbf{W}}_n,\overline{\mathbf{R}}_x)\notag\\
 & +\text{tr}\left\{\nabla_{\mathbf{W}_n}^H N_2(\overline{\mathbf{W}}_n,\overline{\mathbf{R}}_x)(\mathbf{W}_n-\overline{\mathbf{W}}_n)\right\}\notag \\
 & + \text{tr}\left\{\nabla_{\mathbf{R}_x}^H N_2(\overline{\mathbf{W}}_n,\overline{\mathbf{R}}_x)(\mathbf{R}_x-\overline{\mathbf{R}}_x)\right\},
\end{align}
where the gradient of $N_2$ is given by
\begin{align}
  \nabla_{\mathbf{R}_n}^H(\mathbf{W}_n,\mathbf{R}_x) = & -\nabla_{\mathbf{W}_n}^H(\mathbf{W}_n,\mathbf{R}_x) \notag\\
  = & \frac{1}{\ln2} \frac{\widetilde{\mathbf{H}}_n}{\sigma_n^2 + \text{tr}\left\{\widetilde{\mathbf{H}}_n(\mathbf{R}_x-\mathbf{W}_n)\right\}}.
\end{align}

Similarly, with the given $\overline{\mathbf{R}}_x$,
the upper bound of function $n_1$ can be expressed as
\begin{equation}
  n_1(\mathbf{R}_x) \leq \widetilde{n}_1(\mathbf{R}_x).
\end{equation}

The upper bound $\widetilde{n}_1(\mathbf{R}_x)$ is given by
\begin{align}
  \widetilde{n}_1(\mathbf{R}_x) 
 = n_1(\overline{\mathbf{R}}_x)+\text{tr}\left\{\nabla_{\mathbf{R}_x}^H n_1(\overline{\mathbf{R}}_x)(\mathbf{R}_x-\overline{\mathbf{R}}_x)\right\},
\end{align}
where $ \nabla_{\mathbf{R}_x}^H n_1(\mathbf{R}_x)
  = \frac{1}{\ln2} \frac{\widetilde{\mathbf{h}}_e}{\sigma_e^2 + \text{tr}\left\{\widetilde{\mathbf{h}}_e\mathbf{R}_x\right\}}$.

\textcolor{blue}{
Relaxing the rank-one constraint through $\text{rank}(\mathbf{W}_n)=1$, 
the problem (\ref{49}) becomes a convex semidefinite program (SDP), 
which can be efficiently solved to global optimality using standard solvers such as CVX.
Let $\{\tilde{\mathbf{W}}_n\}_{n\in\mathcal{N}}$ and $\tilde{\mathbf{R}}_x$ denote an optimal solution of this SDP.
}

\textcolor{blue}{
Although the relaxed solution $\tilde{\mathbf{W}}_n$ may not be rank-one, 
for the specific structure of our problem, 
a rank-one optimal solution to the original problem (\ref{45}) can always be reconstructed. 
This is a direct consequence of the well-known rank-one recovery property for SDR in downlink beamforming problems with SINR constraint Theorem 1 in \cite{ref38}. 
Specifically, given $\tilde{\mathbf{W}}_n$, the variabel $\mathbf{R}_x$ and $\mathbf{w}_n$ can be constructed as follows:
\begin{align}
    \mathbf{R}_x^* &= \tilde{\mathbf{R}}_x, \label{eq:Rx_opt} \\
    \mathbf{w}_n^* &= (\tilde{\mathbf{h}}_n^H \tilde{\mathbf{W}}_n \tilde{\mathbf{h}}_n)^{-1/2} \tilde{\mathbf{W}}_n \tilde{\mathbf{h}}_n, \quad \forall n \in \mathcal{N}. \label{eq:wn_opt}
\end{align}
}
\textcolor{blue}{
One can verify that $\{\mathbf{w}_n^*\}$ achieves the same objective value as the SDR solution and satisfies all constraints of (\ref{45}).
Hence, it is a global optimum of the original problem.
The optimal covariance matrix of the sensing signal is obtained as:
\begin{equation}
    \mathbf{R}_s^* = \mathbf{R}_x^* - \sum_{n \in \mathcal{N}} \mathbf{w}_n^* (\mathbf{w}_n^*)^H. \label{eq:Rs_opt}
\end{equation}
}

With fixed $\left\{\mathbf{W}, \mathbf{R}_s\right\}$,
the problem $\mathcal{P}1.2$ is formulated as
\begin{align}
  \mathcal{P}1.2: & \max_{\mathbf{U}, \mathbf{F}}
  -\frac{1-\mu}{q_c}\text{tr}(\mathbf{U}^{-1}), \label{58}  \\
  \text{s.t.} \quad & \mathbf{F} = \mathbf{\Theta}_r \mathbf{G} \mathbf{R}_x \mathbf{G}^H \mathbf{\Theta}_r^H, \tag{58a}  \\
  & \begin{bmatrix}
    \mathbf{J}_{\boldsymbol{\theta}\boldsymbol{\theta}}(\mathbf{F}) - \mathbf{U} & \mathbf{J}_{\boldsymbol{\theta}\tilde{\boldsymbol{\alpha}}}(\mathbf{F}) \\
    \mathbf{J}_{\boldsymbol{\theta}\tilde{\boldsymbol{\alpha}}}^T(\mathbf{F}) & \mathbf{J}_{\tilde{\boldsymbol{\alpha}}\tilde{\boldsymbol{\alpha}}}(\mathbf{F})
  \end{bmatrix} \succeq 0. \tag{58b} 
\end{align}

\textcolor{blue}{
Obviously, the maximization of $-\mathrm{tr}(\mathbf{U}^{-1})$ is equivalent to the minimization of $\mathrm{tr}(\mathbf{U}^{-1})$. 
The problem $\mathcal{P}1.2$ can be reformulated as the following problem
\begin{align}
  \mathcal{P}1.2.1: & \min_{\mathbf{U}, \mathbf{F}}
  \frac{1-\mu}{q_c}\text{tr}(\mathbf{U}^{-1}), \label{58}  \\
  \text{s.t.} \quad & \mathbf{F} = \mathbf{\Theta}_r \mathbf{G} \mathbf{R}_x \mathbf{G}^H \mathbf{\Theta}_r^H, \tag{59a} \label{59a} \\
  & \begin{bmatrix}
    \mathbf{J}_{\boldsymbol{\theta}\boldsymbol{\theta}}(\mathbf{F}) - \mathbf{U} & \mathbf{J}_{\boldsymbol{\theta}\tilde{\boldsymbol{\alpha}}}(\mathbf{F}) \\
    \mathbf{J}_{\boldsymbol{\theta}\tilde{\boldsymbol{\alpha}}}^T(\mathbf{F}) & \mathbf{J}_{\tilde{\boldsymbol{\alpha}}\tilde{\boldsymbol{\alpha}}}(\mathbf{F})
  \end{bmatrix} \succeq 0. \tag{59b} 
\end{align}
}

A PDD algorithm is proposed for tackling \textcolor{blue}{$\mathcal{P}1.2.1$}.
\textcolor{blue}{The main idea is to relax the equality constraint (\ref{59a}) using a penalty term in the augmented Lagrangian (AL) framework.}
This penalty term ensures that the solutions of the subproblems remain close to feasible solutions of the original problem.
The algorithm iteratively solves the proposed dual problem, 
while updating the dual variables and penalty terms.
This iterative process continues until convergence is achieved.
Generally speaking, the PDD algorithm operates through a structure of two loops.
Within this structure, Lagrangian dual variables and penalty terms are refined in the outer loop,
and the AL subproblem is addressed in the inner loop.
\textcolor{blue}{For a thorough analysis of the convergence and optimality properties of the PDD framework, readers are referred to \cite{ref36}}

\begin{algorithm}[H]
  \caption{Proposed PDD Algorithm of Transmission Beamforming Optimization}\label{alg:alg1}
  \begin{algorithmic}[1]
    \STATE \textbf{Initialize} all variables $\bm{\chi}^{[0]}$, $\bm{\Upsilon}^{[0]}$, $\rho^{[0]} > 0$,
    $0 < m < 1$, $\rho > 0$, $k = 1$, and threshold $\epsilon$.
    \STATE \textbf{Outer Loop}: 
    \REPEAT
      \STATE \textbf{Inner Loop}:
      \STATE $\quad \bm{\chi}^{[k+1]} = \text{optimize}(\mathcal{P}_{\text{AL}}(\rho^{[k]}, \bm{\Upsilon}^{[k]})),$
        \IF{$h(\bm{\chi}^{[k+1]}) \leq \eta^{[k]},$}
          \STATE $\quad \bm{\Upsilon}^{[n+1]} = \bm{\Upsilon}^{[k]} + \frac{1}{\rho^{[k]}}(\mathbf{F}^{[k+1]} - \bm{\Theta}_r^{[n+1]} \mathbf{G} \mathbf{R}_x^{[n+1]} \times \mathbf{G}^H (\bm{\Theta}_r^{[n+1]})^H),$
          \STATE $\quad \rho^{[k+1]} = \rho^{[k]}.$
          \ELSE
          \STATE $\quad \bm{\Upsilon}^{[k+1]} = \bm{\Upsilon}^{[k]}$, $\rho^{[k+1]} = m\rho^{[k]}.$
          \ENDIF
        \STATE \quad $k = k + 1.$
      \UNTIL{the constraint violation function $g(\bm{\chi})$ drops beneath the threshold $\epsilon$.}
  \end{algorithmic}
\end{algorithm}

To derive the Lagrange dual variable matrix \(\boldsymbol{\Upsilon} \in \mathbb{C}^{K \times K}\)
and the penalty parameter \(\rho > 0\),
associated with a constraint \(\mathbf{F} = \boldsymbol{\Theta}_r \mathbf{G} \mathbf{R}_x \mathbf{G}^H \boldsymbol{\Theta}_r^H\), 
AL formulation of the subproblem \textcolor{blue}{\(\mathcal{P}1.2.1\)} is derived by
\begin{align}
  \mathcal{P}_{\text{AL}}:
  \min_{\mathbf{U}, \mathbf{F}}
  & \frac{1-\mu}{q_c}\left[\text{tr}(\mathbf{U}^{-1}) + \mathcal{P}^{\rho}(\bm{\chi}, \bm{\Upsilon})\right], \label{59}\\
  \text{s.t.} \quad & \begin{bmatrix}
    \mathbf{J}_{\boldsymbol{\theta}\boldsymbol{\theta}}(\mathbf{F}) - \mathbf{U} & \mathbf{J}_{\boldsymbol{\theta}\tilde{\boldsymbol{\alpha}}}(\mathbf{F}) \\
    \mathbf{J}_{\boldsymbol{\theta}\tilde{\boldsymbol{\alpha}}}^T(\mathbf{F}) & \mathbf{J}_{\tilde{\boldsymbol{\alpha}}\tilde{\boldsymbol{\alpha}}}(\mathbf{F})
  \end{bmatrix} \succeq 0, \tag{60a} 
\end{align}
where 
$\mathcal{P}^{\rho}(\boldsymbol{\chi}, \boldsymbol{\Upsilon}) = \frac{1}{2\rho} \left\| \mathbf{F} - \boldsymbol{\Theta}_r \mathbf{G} \mathbf{R}_x \mathbf{G}^H \boldsymbol{\Theta}_r^H + \rho \boldsymbol{\Upsilon} \right\|_F^2$.
\textcolor{blue}{The expression $\|\cdot\|_F$ denotes the Frobenius norm}.
The constraint violation is defined by the following function
\begin{equation}
g(\boldsymbol{\chi}) = \|\mathbf{F} - \boldsymbol{\Theta}_r \mathbf{G} \mathbf{R}_x \mathbf{G}^H \boldsymbol{\Theta}_r^H\|_\infty. 
\end{equation}
A vanishing sequence $\{\eta^{[n]}\}_{n=1}^\infty$ is considered where $\eta^{[n]} = 0.99g(\boldsymbol{\chi}^{[n-1]})$. 
For a small penalty parameter $\rho$, there will be \(\lim_{n\to\infty} \mathcal{P}^\rho(\boldsymbol{\chi}^{[n]}) = 0 \quad \text{and} \quad \lim_{n\to\infty} g(\boldsymbol{\chi}^{[n]}) = 0\), 
which satisfies the constraint (\ref{43a}).
\textcolor{blue}{For fixed values of $\boldsymbol{\Theta}_r$, $\mathbf{R}_x$, $\rho$, and $\boldsymbol{\Upsilon}$ in each inner iteration, the AL subproblem $\mathcal{P}_{\text{AL}}$ is a convex optimization problem. This is because the objective function is a linear combination of the convex function $\mathrm{tr}(\mathbf{U}^{-1})$ and the convex quadratic penalty term $\mathcal{P}^{\rho}$, subject to a linear matrix inequality constraint. Therefore, $\mathcal{P}_{\text{AL}}$ can be efficiently solved via CVX in the inner loop of the PDD algorithm.}

\subsubsection{\textbf{Optimization of STAR-RIS phase shifts}}
By fixing $\mathbf{U}$, $\mathbf{W}$ and $\mathbf{R}_s$,
the STAR-RIS phase shifts problem is expressed as
\begin{align}
  \mathcal{P}2: \max_{\mathbf{\bm{\theta}}_t, \mathbf{\bm{\theta}}_r}
  & \frac{\mu}{q_s}\sum_{n\in\mathcal{N}}(\text{R}_n-\text{R}_e) + 
  \frac{\mu-1}{q_c} \text{tr}(\mathbf{U}^{-1}), \label{61}\\
  \text{s.t.}
  & \beta_{t,k}^2 + \beta_{r,k}^2 = 1, 0 \leq \beta_{t,k}, \beta_{r,k} \leq 1, \forall k. \tag{61a} \label{61a}
\end{align}

The constraint (\ref{61a}) is expressed as
\begin{equation}
  |\boldsymbol{\theta}_t[k]|^2 + |\boldsymbol{\theta}_r[k]|^2 = 1. \label{62}
\end{equation}

To solve the subproblem \(\mathcal{P}2\), the projected gradient method is applied.

\textcolor{blue}{
Since the objective function is real-valued but depends on complex variables $\boldsymbol{\Theta}_t$ and $\boldsymbol{\Theta}_r$, 
Wirtinger calculus is employed for differentiation. 
$\boldsymbol{\Theta}_t$ and $\boldsymbol{\Theta}_t^*$ are treated as independent variables. 
The gradient used for updating $\boldsymbol{\Theta}_t$ in the descent direction is the derivative with respect to $\boldsymbol{\Theta}_t^*$, 
denoted as $\nabla_{\boldsymbol{\Theta}_t^*} f$. 
Similarly, $\nabla_{\boldsymbol{\Theta}_r^*} f$ is used for updating $\boldsymbol{\Theta}_r$.
}

By defining the auxiliary variables 
${\hat{\mathbf{R}}_x} = \mathbf{G}\mathbf{R}_x\mathbf{G}^H$,
${\widetilde{\mathbf{h}}_n} = \mathbf{h}_n\mathbf{h}_n^H$
$ \mathbf{M}=\mathbf{g}_n\mathbf{h}_n^H + \mathbf{G}^H\mathbf{\Theta}_t^H \widetilde{\mathbf{h}}_n$,
and $ \mathbf{N}_n = \mathbf{h}_n\mathbf{g}_n^H(\mathbf{R}_x - \mathbf{W}_n)\mathbf{G}^H + \widetilde{\mathbf{h}}_n \mathbf{\Theta}_t\mathbf{G} (\mathbf{R}_x - \mathbf{W}_n)\mathbf{G}^H$,
\textcolor{blue}{
the partial derivative $\frac{\partial f}{\partial \boldsymbol{\Theta}_t}$ and
the gradient of $\mathbf{\Theta}_t^*$ are given by
}
\begin{align}
\textcolor{blue}{\frac{\partial f}{\partial \boldsymbol{\Theta}_t}} = & \frac{\mu}{\ln2 q_s}
\sum_{n\in\mathcal{N}}\frac{2\left({\mathbf{G} \mathbf{R}_x \mathbf{g}_n \mathbf{h}_n^H + \hat{\mathbf{R}}_x \mathbf{\Theta}_t^H {\widetilde{\mathbf{h}}_n}}\right)^T}
  {\sigma_n^2 + \mathrm{tr} \left\{ \mathbf{\widetilde{H}}_n\mathbf{R}_x \right\}} \notag\\
  & - \frac{\mu}{\ln2 q_s} \sum_{n\in\mathcal{N}}\frac{2\left(\mathbf{G} (\mathbf{R}_x - \mathbf{W}_n) \mathbf{M}\right)^T}
  {\sigma_n^2 + \mathrm{tr} \left\{\mathbf{\widetilde{H}}_n(\mathbf{R}_x - \mathbf{W}_n) \right\}},
\end{align}
\begin{align}
  \nabla_{\mathbf{\Theta}_t^*} f = & \frac{\mu}{\ln2 q_s}
  \sum_{n\in\mathcal{N}}\frac{2\left(\mathbf{h}_n \mathbf{g}_n^H \mathbf{R}_x \mathbf{G}^H +  \widetilde{\mathbf{h}}_n  \mathbf{\Theta}_t \hat{\mathbf{R}}_x \right)}
    {\sigma_n^2 + \mathrm{tr} \left\{ \mathbf{\widetilde{H}}_n\mathbf{R}_x \right\}} \notag\\
    & - \frac{\mu}{\ln2 q_s} \sum_{n\in\mathcal{N}}\frac{2 \mathbf{N}_n}
    {\sigma_n^2 + \mathrm{tr} \left\{\mathbf{\widetilde{H}}_n(\mathbf{R}_x - \mathbf{W}_n) \right\}}.
\end{align}

\textcolor{blue}{
The partial derivative $\frac{\partial f}{\partial \boldsymbol{\Theta}_r}$ and
the gradient of $\mathbf{\Theta}_r^*$ are given by
}
\begin{align}
  \textcolor{blue}{\frac{\partial f}{\partial \boldsymbol{\Theta}_t}} =&  -\frac{2\mu}{\ln2 q_s}
  \sum_{n\in\mathcal{N}}\frac{(\hat{\mathbf{R}}_x \mathbf{\Theta}_r^H \mathbf{h}_e \mathbf{h}_e^H)^T}
  {\sigma_e^2 + \text{tr} \left\{ \widetilde{\mathbf{h}}_e \mathbf{R}_x\right\}}\notag\\
  & + \frac{2\mu}{\ln2 q_s} \sum_{n\in\mathcal{N}}\frac{( \mathbf{h}_e \mathbf{h}_e^H \mathbf{\Theta}_r \mathbf{G} (\mathbf{R}_x - \mathbf{W}_n) \mathbf{G}^H)^T}
  {\sigma_e^2 + \text{tr} \left\{ \widetilde{\mathbf{h}}_e (\mathbf{R}_x - \mathbf{W}_n)\right\}},
\end{align}

\begin{align}
  \nabla_{\mathbf{\Theta}_r^*}f =&  -\frac{2\mu}{\ln2 q_s}
  \sum_{n\in\mathcal{N}}\frac{\mathbf{h}_e \mathbf{h}_e^H \mathbf{\Theta}_r \hat{\mathbf{R}}_x}
  {\sigma_e^2 + \text{tr} \left\{ \widetilde{\mathbf{h}}_e \mathbf{R}_x\right\}}\notag\\
  & + \frac{2\mu}{\ln2 q_s} \sum_{n\in\mathcal{N}} \frac{\mathbf{h}_e \mathbf{h}_e^H \mathbf{\Theta}_r \mathbf{G} (\mathbf{R}_x - \mathbf{W}_n) \mathbf{G}^H}
  {\sigma_e^2 + \text{tr} \left\{ \widetilde{\mathbf{h}}_e (\mathbf{R}_x - \mathbf{W}_n) \right\}}.
\end{align}

The vectors $\boldsymbol{\theta}_t$ and $\boldsymbol{\theta}_r$ are composed of diagonal elements of the matrix $\mathbf{\Theta}_t$ and $\mathbf{\Theta}_r$, respectively.
The feasible set can be described as: ${\Theta} = \left\{\boldsymbol{\theta}_t, \boldsymbol{\theta}_r\,|\;|\boldsymbol{\theta}_t[k]|^2 + |\boldsymbol{\theta}_r[k]|^2 =1\right\}$.
The entire algorithmic process is detailed in Algorithm 3.
\textcolor{blue}{
The constraint $|\theta_{t,k}|^2 + |\theta_{r,k}|^2 = 1$ for each STAR-RIS element defines a non-convex manifold. 
To enforce this constraint while performing gradient-based optimization, we employ a projected gradient method. 
After each gradient update step, the updated coefficients are projected onto the feasible set via element-wise normalization. 
Specifically, for each element $k$, let $\theta_{t,k}^{\mathrm{temp}}$ and $\theta_{r,k}^{\mathrm{temp}}$ denote the values obtained after the gradient ascent step. 
The projection operation is defined as:
\begin{equation}
\label{eq:projection}
(\theta_{t,k}, \theta_{r,k}) \leftarrow \frac{(\theta_{t,k}^{\mathrm{temp}}, \theta_{r,k}^{\mathrm{temp}})}
{\sqrt{|\theta_{t,k}^{\mathrm{temp}}|^2 + |\theta_{r,k}^{\mathrm{temp}}|^2}}, \quad \forall k \in \mathcal{K}.
\end{equation}
This normalization ensures that $|\theta_{t,k}|^2 + |\theta_{r,k}|^2 = 1$ for every element $k$ while preserving the relative phase between transmission and reflection coefficients. 
The complete projected gradient algorithm, incorporating this projection step, is summarized in Algorithm~\ref{alg:alg2}.
}

\begin{algorithm}
  \caption{Proposed Projected Gradient Algorithm of STAR-RIS Optimization}\label{alg:alg2}
  \begin{algorithmic}[1]
    \STATE \textbf{Input}: $\bm{\Theta}_r^{(0)}$, $\bm{\Theta}_t^{(0)}$, $\mu_r$, $\mu_t$.
    \STATE \textbf{for} $k = 1, 2, ...$ \textbf{do}
    \STATE $\qquad \bm{\Theta}_r^{[k+1]} = \bm{\Theta}_r^{[k]} + \mu_r \nabla_{\bm{\Theta}_r^*} f,$
    \STATE $\qquad \bm{\Theta}_t^{[k+1]} = \bm{\Theta}_t^{[k]} + \mu_t \nabla_{\bm{\Theta}_t^*} f,$
    \STATE $\qquad \bm{\theta}_r^{[k+1]} = \text{diag}(\bm{\Theta}_r^{[k+1]}), \bm{\theta}_t^{[k+1]} = \text{diag}(\bm{\Theta}_t^{[i+1]}), $
    \STATE \textbf{end for}
  \end{algorithmic}
\end{algorithm}

\textcolor{blue}{
The computational complexity of the proposed algorithm primarily arises from iteratively solving two coupled subproblems: 
the transmit beamforming optimization subproblem and the STAR-RIS phase shifts optimization subproblem. 
The transmit beamforming subproblem is reformulated as a SDP problem, solved by the interior-point method within the PDD framework. 
Given a solution accuracy $\varepsilon$, the per-iteration complexity is calculated as $\mathcal{O}\left( (N^{6.5} M^{6.5} + K^{6.5}) \log(1/\varepsilon) \right)$,
where $M$, $N$, and $K$ denote the number of BS antennas, communication users, and STAR‑RIS elements, respectively \cite{ref39}. 
The STAR‑RIS phase shifts subproblem is addressed by the PGM, 
whose per-iteration complexity is dominated by gradient computations, 
costing $\mathcal{O}(K^2 M + K M^2)$ operations.
}

\section{SIMULATION RESULTS} \label{section5}
Simulation results are provieded to demonstrate effectiveness of the proposed joint beamforming and STAR-RIS algorithm in this section.
The point target (eavesdropper) is a far apart scattered point for STAR-RIS, which is located at (30, 120, 30) m.
The BS is positioned at (30, 40, 0) m,
while STAR-RIS is placed at (0, 0, 0) m.
Legitimate communication users are located randomly in the communicating zone.
\textcolor{blue}{The tradeoff weight parameter $\mu$ is set to $0.5$ unless otherwise specified.
In addition, the normalization constants are calculated based on the maximum achievable secrecy rate and the minimum CRB, respectively.
The constant $q_s$ is set to $1$, and $q_c$ is set to $1\text{e}^{-6}$.}
Assume that the communication channel is Rayleigh fading, and the noise is additive white Gaussian noise.
\textcolor{blue}{The main system parameters are given in Table \ref{table1}.
The simulations are implemented in MATLAB R2022b on a computer with Apple M2 Pro chip.
The CVX toolbox 2.2 is used to solve the convex optimization problems.}

\begin{figure}[!t]
  \centering
  \includegraphics[width=3.5in]{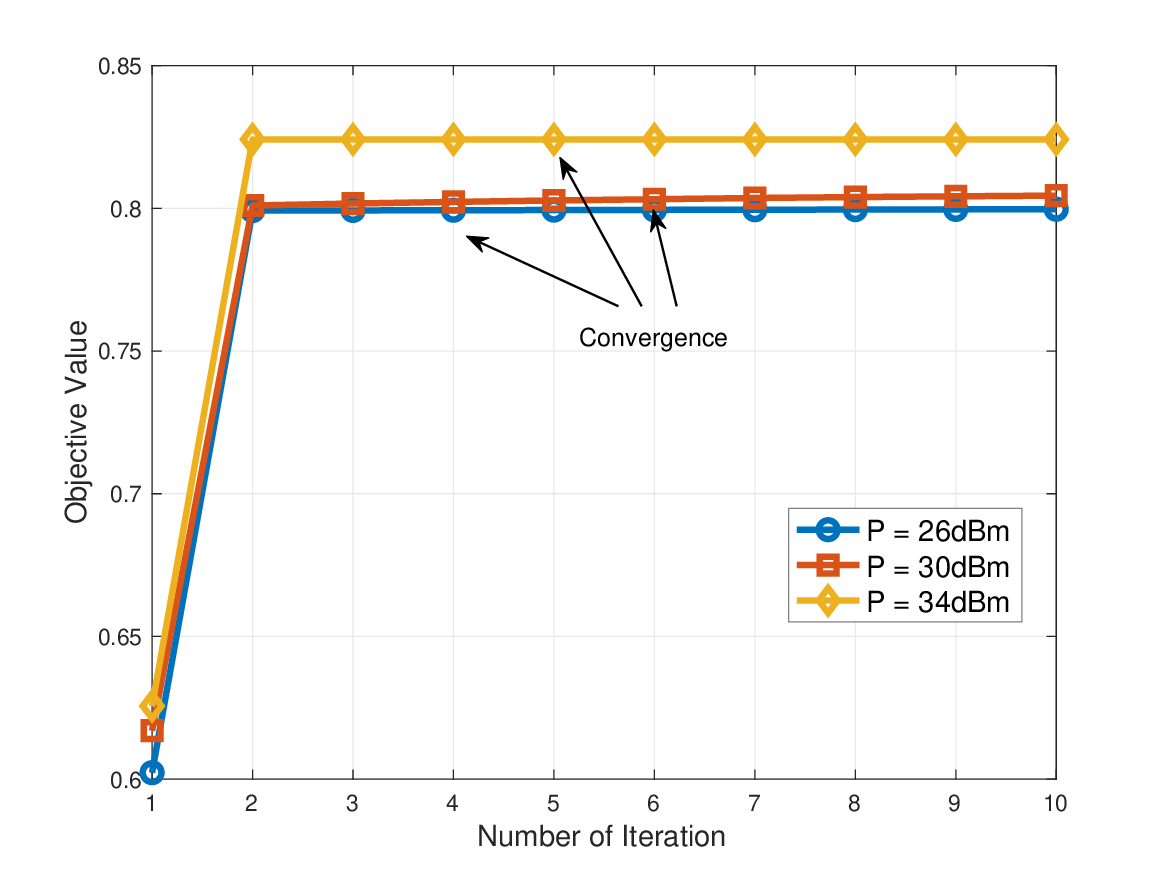}
  \caption{Convergence behavior of algorithm}
  \label{fig_2}
\end{figure}

\textcolor{blue}{
The iterative convergence behavior of the proposed algorithm is illustrated in Fig. 2. 
The objective function value, 
which jointly reflects secrecy rate and CRB is plotted against the iteration index under varying transmit power levels. 
It is observed that the algorithm converges to a stable solution within a small number of iterations, typically 4 to 6, 
demonstrating its computational efficiency and practical feasibility. 
This rapid convergence can be attributed to the well-structured subproblems,
where the transmit beamforming is solved via a convex SDP after SCA relaxation, 
and the STAR-RIS phase shifts are updated via a projected gradient method,
allowing each block to be efficiently optimized while fixing the others. 
As expected, the converged objective value exhibits a positive correlation with the transmit power budget, confirming that higher power improves the joint performance.
}

\begin{figure}[!t]
  \centering
  \includegraphics[width=3.5in]{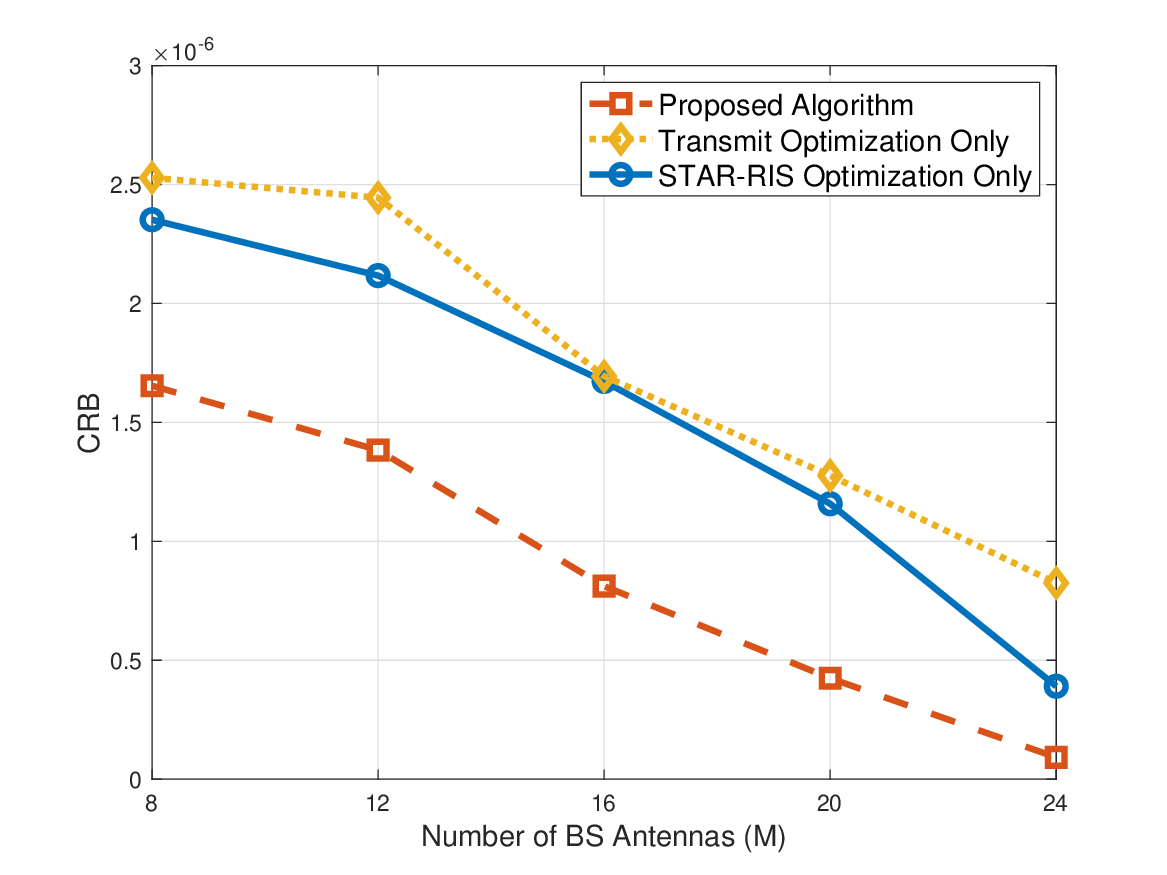}
  \caption{CRB versus BS antennas with two benchmarks}
  \label{fig_8}
\end{figure}

Fig. \ref{fig_8} presents the CRB performance as a function of the number of BS antennas $M$ under three different schemes: 
the proposed hybrid BCD algorithm, transmit optimization only, and STAR-RIS optimization only. 
As the number of BS antennas increases, all schemes demonstrate a decreasing CRB trend,
indicating enhanced estimation accuracy. 
The proposed algorithm consistently achieves the lowest CRB across all antenna configurations, 
clearly outperforming the two benchmark methods. 
Notably, the STAR-RIS optimization scheme provides better performance than the transmit optimization only approach,
underscoring the effectiveness of RIS-based enhancements.
Overall, the results highlight the superior efficiency of the joint optimizating strategy adopted before.

\begin{figure}[!t]
  \centering
  \includegraphics[width=3.5in]{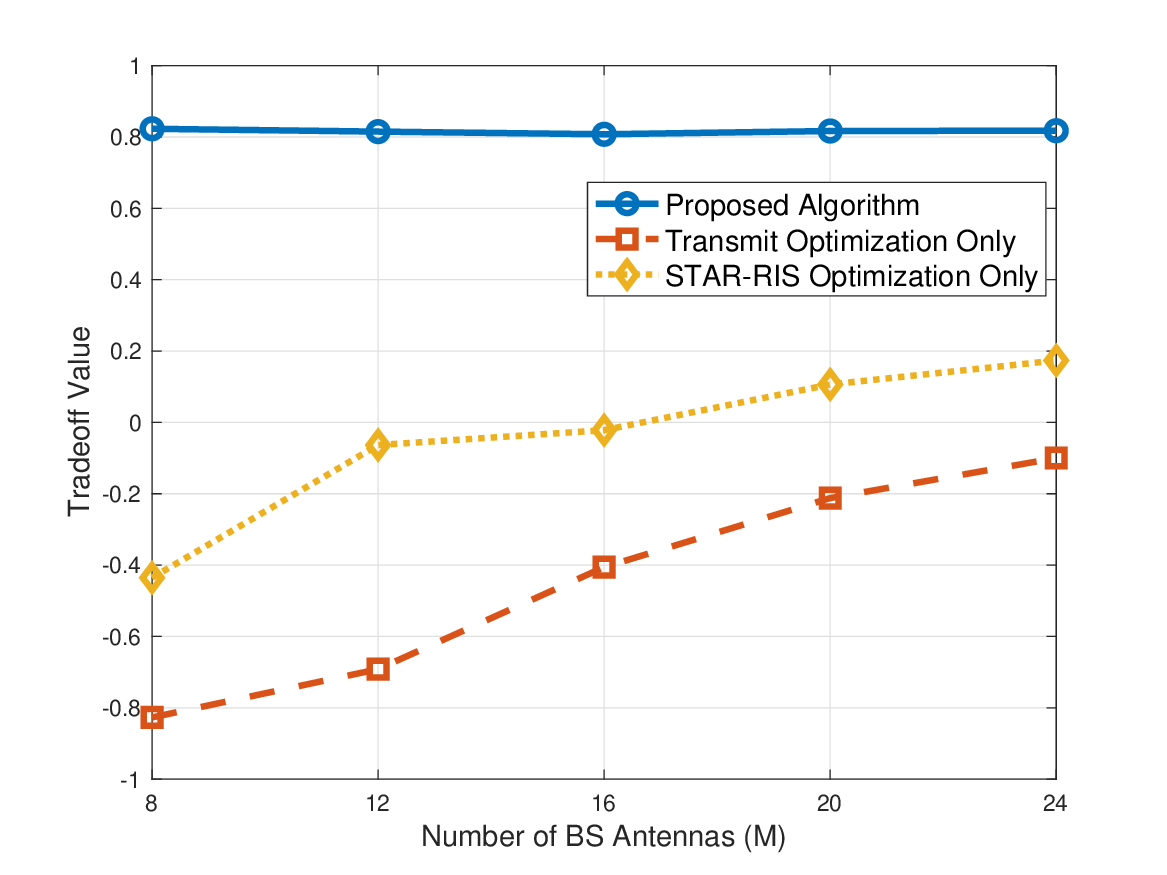}
  \caption{Tradeoff Value versus BS antennas with two benchmarks}
  \label{fig_9}
\end{figure}

Fig. \ref{fig_9} illustrates how the tradeoff performance varies with the number of BS antennas $M$.
The proposed algorithm consistently achieves a high tradeoff value across all antenna configurations,
with the weight $\mu$ held constant, 
demonstrating strong robustness and balanced optimization of communication and sensing capabilities.
In comparison, the transmit-optimization-only algorithm starts with a severely degraded tradeoff value when the number of BS antennas is small. 
Although performance improves steadily with increasing $M$, 
it remains significantly inferior to the proposed approach. 
This suggests that optimizing the transmission variables alone is insufficient to meet sensing requirements, 
particularly in small-scale antenna deployments.
The STAR-RIS-optimization-only approach exhibits moderate performance, 
with tradeoff values gradually shifting as $M$ increases.
\textcolor{blue}{
This trend underscores the fundamental resource competition between the two functions: more antennas provide additional degrees of freedom, but they must be allocated between forming beams for confidential communication and steering energy for accurate eavesdropper sensing. The proposed joint optimization dynamically balances this allocation, whereas the benchmarks suffer because they only optimize one side of this tradeoff.
}
However, both benchmark schemes are clearly outperformed by the proposed joint optimization, 
underlining the necessity of coordinated transmitter and STAR-RIS control in ISAC systems.

\begin{figure}[!t]
  \centering
  \includegraphics[width=3.5in]{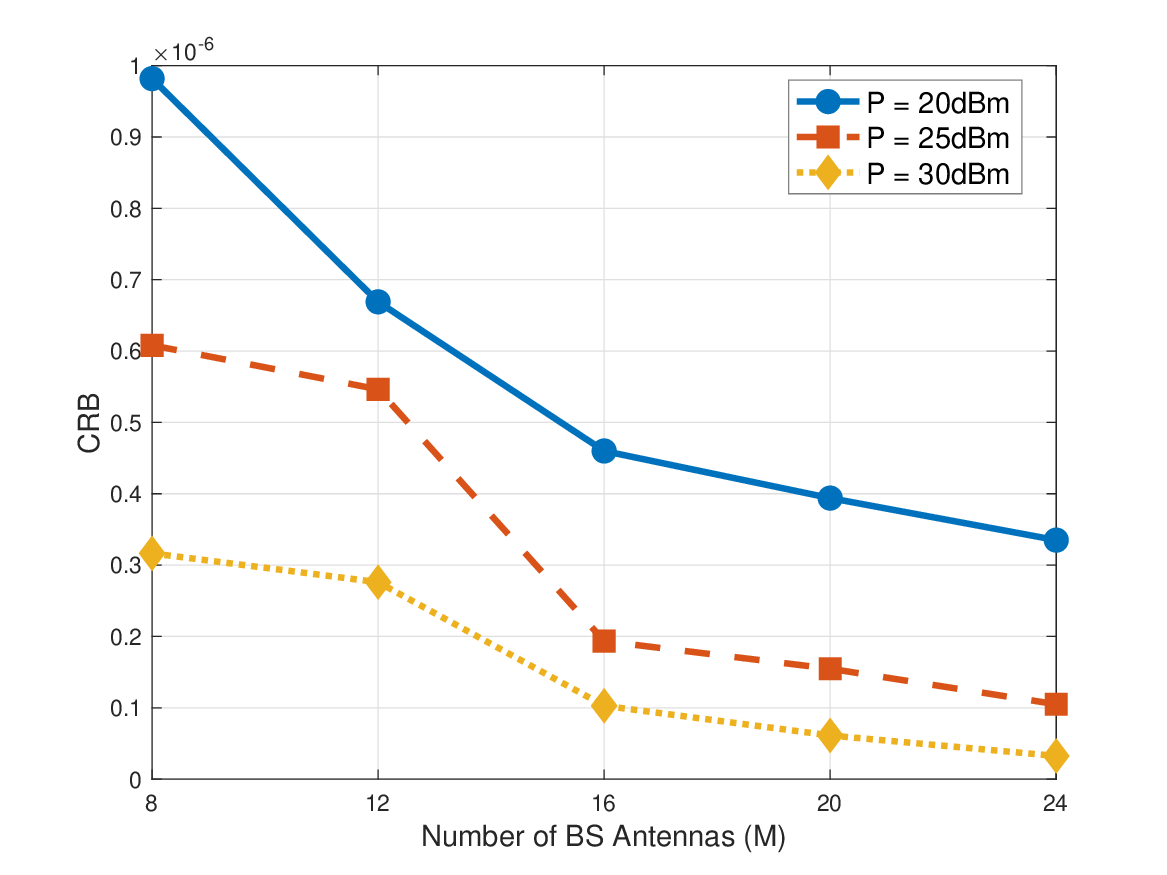}
  \caption{CRB versus different BS antennas}
  \label{fig_3}
\end{figure}

\begin{table}
\renewcommand{\arraystretch}{1.1}
\caption{System Parameters} \label{table1}
\centering
\begin{tabular}{|c|c|}
\hline
Number of BS antennas $M$ & 16 \\
\hline
Number of STAR-RIS elements $K$ & 8 \\
\hline
Number of sensors $S$ & 8 \\
\hline
Number of communication users $N$ & 4 \\
\hline
The noise power $\sigma_n^2, \sigma_e^2$ & -110dBm\\
\hline
Transmit power at the BS $P$ & 30dBm \\
\hline

\end{tabular}
\end{table}
Fig. \ref{fig_3} depicts CRB verus the number of BS antennas under varying transmit power levels. 
As observed, CRB exhibits a consistent decline with increasing BS antennas across differnet power levels, 
demonstrating the enhancement of the eavesdropper estimation accuracy.
This is because the larger transmsion beamforming gain and the more beamforming towards STAR-RIS
thereby improve overall sensing capability.
In addition, the increase of transmit power results in the reduction of CRB.
The underlying reason is that enhanced transmit power boosts the eavesdropper’s SNR, 
resulting in more accurate eavesdropper estimation.

\begin{figure}[!t]
  \centering
  \includegraphics[width=3.5in]{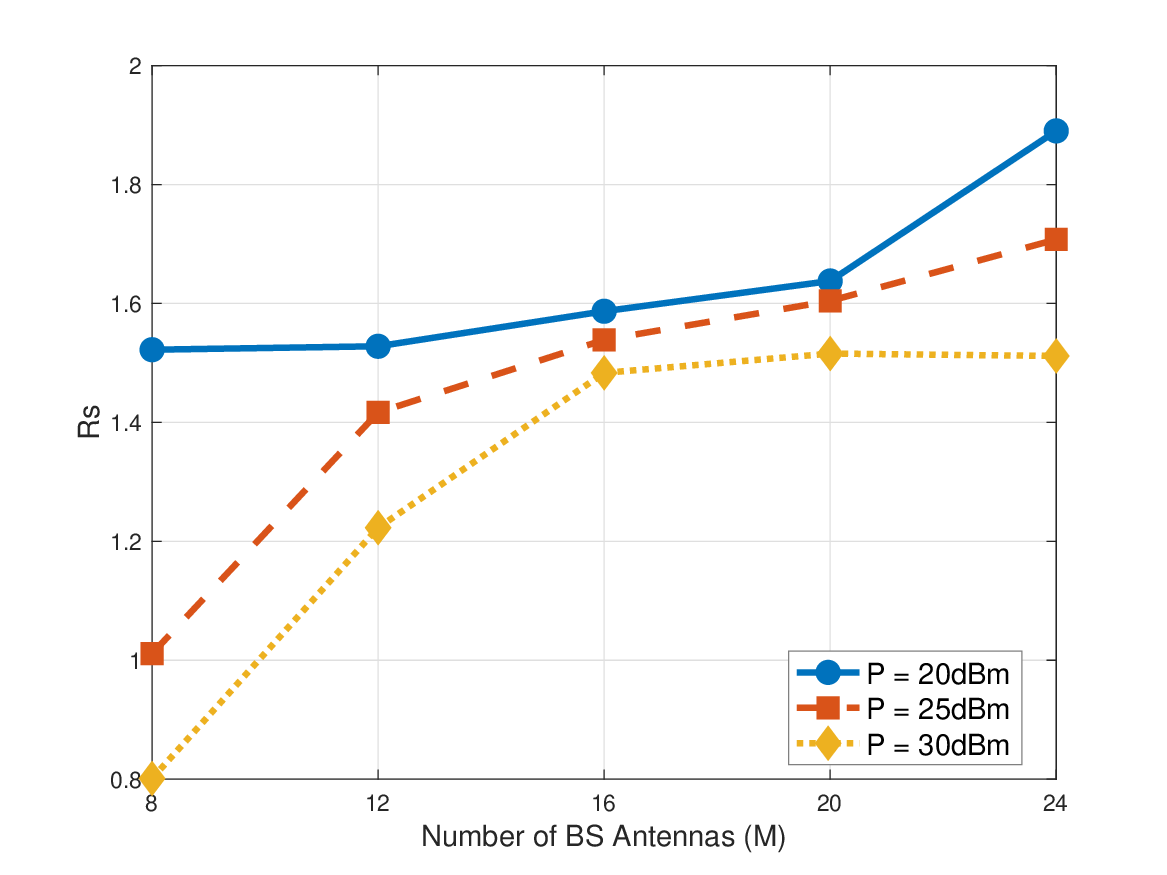}
  \caption{Secrecy Rate versus different BS antennas}
  \label{fig_4}
\end{figure}

Fig. \ref{fig_4} illustrates the \textcolor{blue}{secrecy rate (in bit/s/Hz)} with various BS \textcolor{blue}{antennas} under different transmit power levels, 
As the number of BS anntennas grows,
the secrecy rate consistently increases across all transmit power levels,
indicating that less information is leaken to the eavesdropper.
This is because more BS antennas provide the larger transmit beamforming gain and the more beamforming towards STAR-RIS.

\begin{figure}[!t]
  \centering
  \includegraphics[width=3.5in]{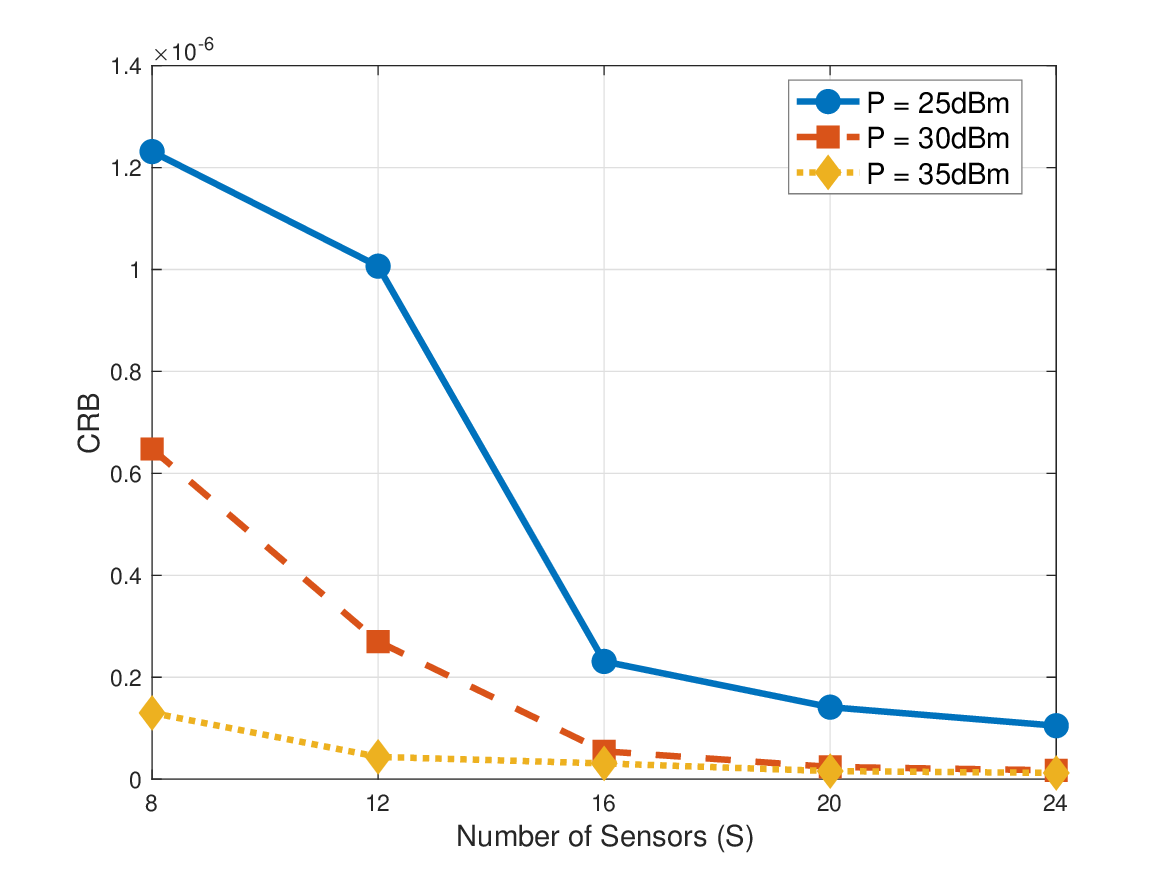}
  \caption{CRB versus different sensors}
  \label{fig_6}
\end{figure}
Fig. \ref{fig_6} shows the eavesdropper sensing performance versus different STAR-RIS elements and sensors.
Obviously, the value of CRB decreases while sensors increase.
The reduction in CRB arises from the growth in the number of sensors, 
which leads to a higher-rank FIM. 
This increase in rank reflects richer and more diverse measurements, 
ultimately improving estimation performance.
The sensing performance improves with more STAR-RIS elements, 
as they offer greater degrees of freedom for wavefront shaping, 
resulting in enhanced observation quality and lower estimation error.

\begin{figure}[!t]
  \centering
  \includegraphics[width=3.5in]{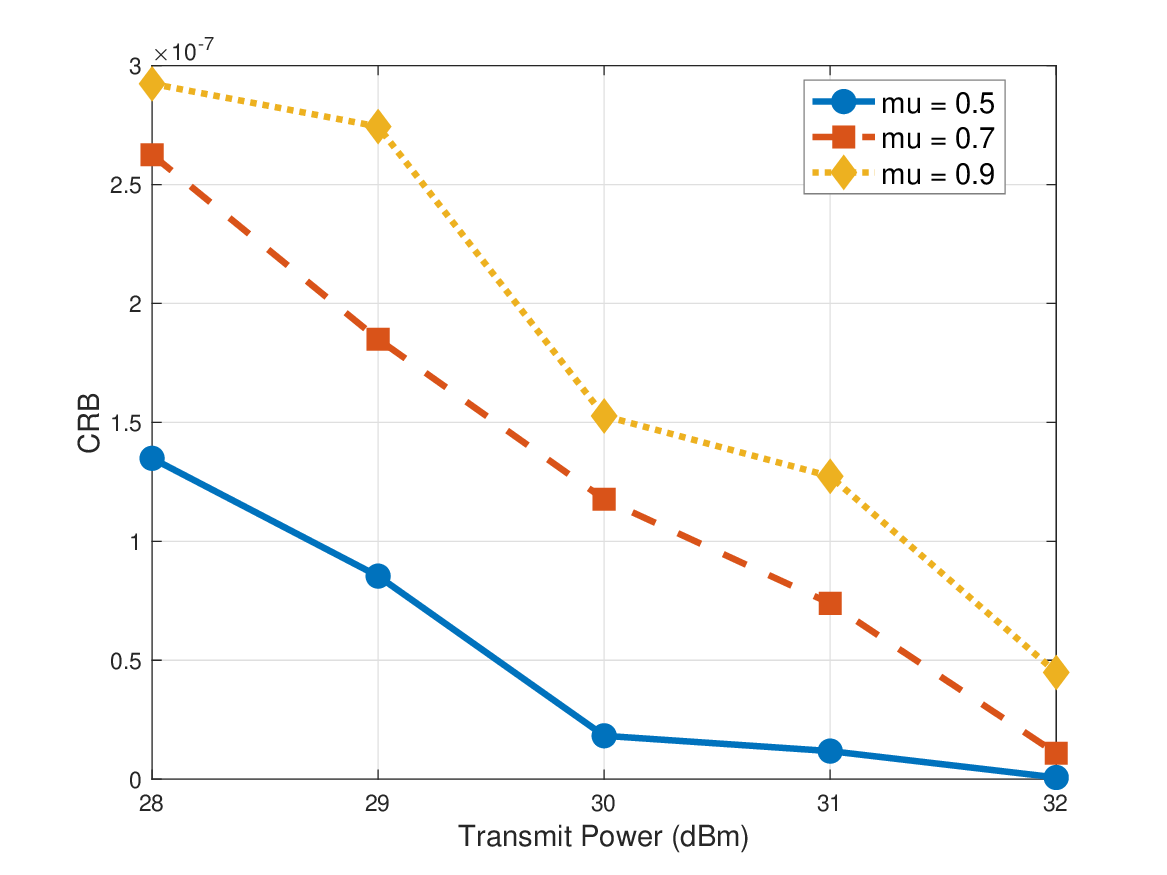}
  \caption{CRB versus different transmit power at different tradeoff weights}
  \label{fig_7}
\end{figure}
Fig. \ref{fig_7} shows the CRB behavior in DOA estimation with respect to transmit power under three different tradeoff weights $\mu$. 
The CRB decreases monotonically with increasing transmit power,
demonstrating that higher transmit power improves sensing accuracy by enhancing the SNR at the sensors. 
However, increasing the tradeoff weight, 
which prioritizes secrecy rate over sensing performance,
leads to higher CRB value across all power levels. 
This results \textcolor{blue}{in} the inherent tradeoff between communication security and estimation precision. 
The result indicates that while higher transmit power enhances sensing performance,
the choice of $\mu$ significantly impacts the achievable CRB.

\section{CONCLUSION} \label{section6}
A STAR-RIS-empowered ISAC system with an eavesdropper was proposed to enhance the security and the sensing performance in this paper.
First, he CRB of eavesdropper estimation was derived, and the secrecy rate was formulated to improve security of the system.
Next, the optimization problem for STAR-RIS phase shifts and transmit beamforming was put forward, 
seeking an optimal tradeoff between sensing accuracy and communication security,
which was reformulated as a non-convex optimization,
which was solved by the proposed hybrid BCD algorithm.
Variables were decomposed into three parts.
For the transmit beamforming optimization, 
a hybrid algorithm integrating the PDD framework was proposed with SCA techniques. 
Regarding the STAR-RIS phase shifts optimization, 
the projected gradient method was implemented to achieve optimal solutions.
Simulation results revealed the substantial efficiency of the algorithm put forward,
which effectively improved accurate eavesdropper sensing and maintained a high secrecy rate, highlighting its strong tradeoff capabilities.
\textcolor{blue}{
Possible extensions of this work include studying multi-target sensing scenarios, considering dynamic environments with mobile targets/users, investigating the impact of imperfect STAR-RIS hardware, 
and exploring data-driven optimization methods for large-scale deployments. 
Integrating the proposed framework with emerging 6G technologies, such as reconfigurable holographic surfaces and joint communication and sensing waveforms, also presents promising research avenues.
}

\renewcommand{\refname}{REFERENCES}

\begin{IEEEbiography}[{\includegraphics[width=1in,height=1.25in,clip,keepaspectratio]{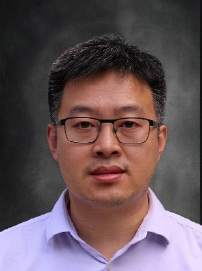}}]{Haijun Zhang} (Fellow, IEEE) is currently a Full Professor at University of Science and Technology Beijing, China. He was a Postdoctoral Research Fellow in Department of Electrical and Computer Engineering, the University of British Columbia (UBC), Canada. He serves/served as Track Co-Chair of VTC Fall 2022 and WCNC 2020/2021, Symposium Chair of Globecom’19, TPC Co-Chair of INFOCOM 2018 Workshop on Integrating Edge Computing, Caching, and Offloading in Next Generation Networks, and General Co-Chair of GameNets’16. He serves as an Editor of IEEE Transactions on Wireless Communications, IEEE Transactions on Information Forensics and Security. He received the IEEE CSIM Technical Committee Best Journal Paper Award in 2018, IEEE ComSoc Young Author Best Paper Award in 2017, IEEE ComSoc Asia-Pacific Best Young Researcher Award in 2019. He is a Distinguished Lecturer of IEEE and IEEE Fellow.
\end{IEEEbiography}
\begin{IEEEbiography}[{\includegraphics[width=1in,height=1.25in,clip,keepaspectratio]{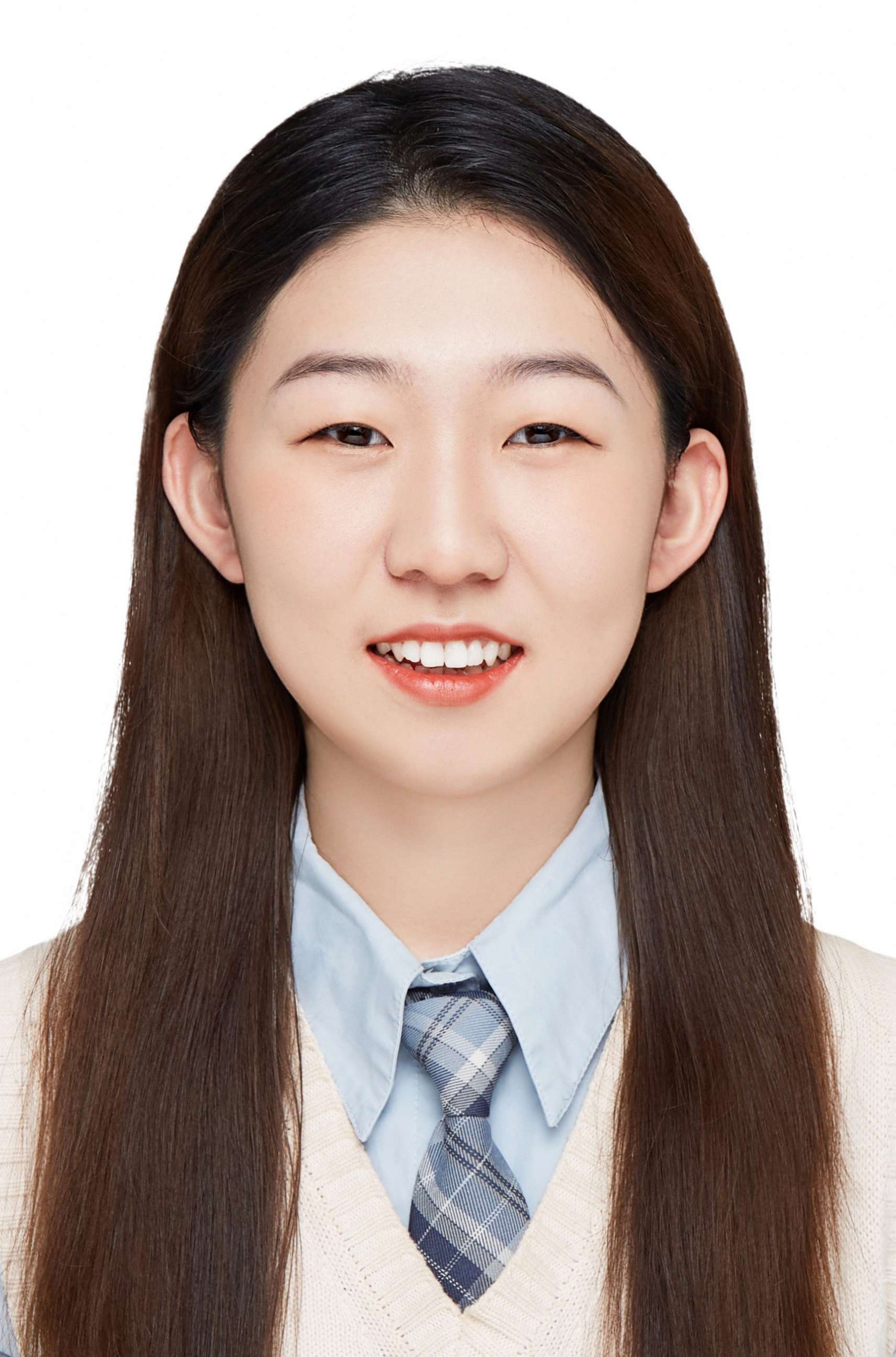}}]{Shuqing Wu} received the B.S. degree from the School of Computer and Communication Engineering, University of Science and Technology Beijing, Beijing, China, in 2023, where she is currently pursuing the M.S. degree. Her research interest is integrated sensing and communications.
\end{IEEEbiography}
\begin{IEEEbiography}[{\includegraphics[width=1in,height=1.25in,clip,keepaspectratio]{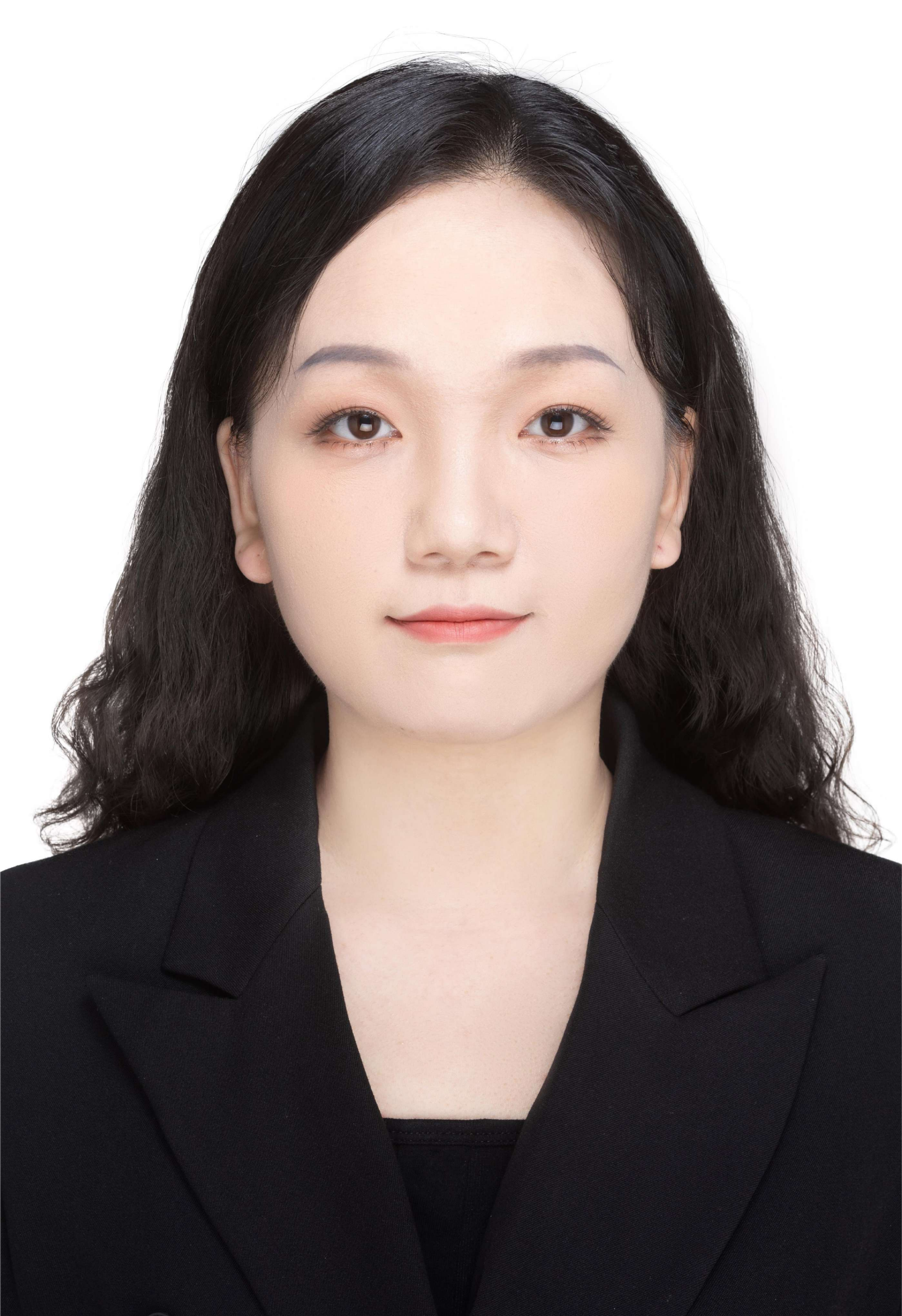}}]{Xiaoqi Zhang} (Member, IEEE) received the Ph.D. degree from the University of Science and Technology Beijing, Beijing, China, in 2024. She was a Postdoctoral Fellow with The Hong Kong Polytechnic University, Hong Kong. She is currently an Associate Professor with the University of Science and Technology Beijing. Her research interests include intelligent reflecting surfaces, integrated sensing and communication, resource allocation with UAV communications, and next generation wireless communication.
\end{IEEEbiography}
\begin{IEEEbiography}[{\includegraphics[width=1in,height=1.25in,clip,keepaspectratio]{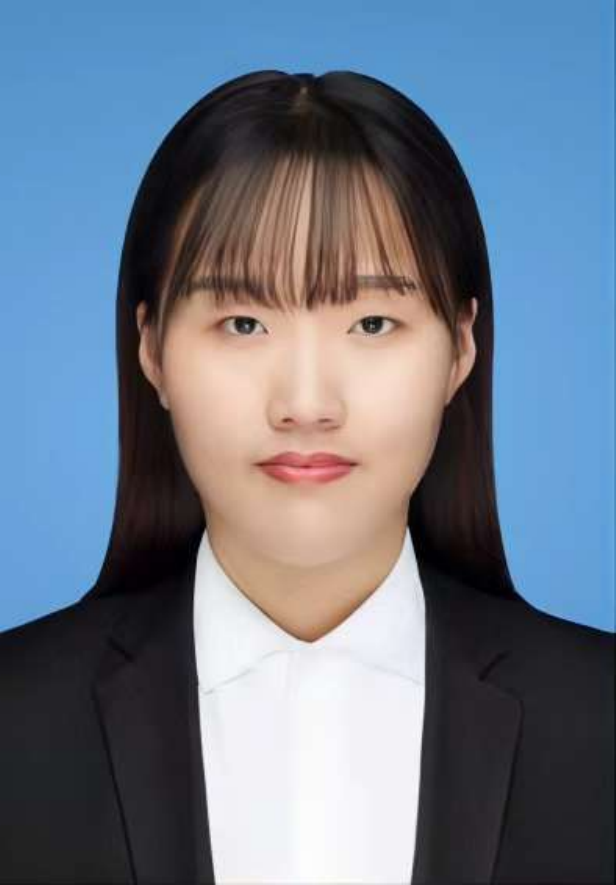}}]{Zijun Wu} received the B.S. degree from the School of Computer and Communication Engineering, University of Science and Technology of Beijing, Beijing, China, in 2021, where she is currently pursuing the Ph.D. degree. Her research interests include IRS, mobile edge computing, and resource allocation in 6G wireless communication.
\end{IEEEbiography}
\begin{IEEEbiography}[{\includegraphics[width=1in,height=1.25in,clip,keepaspectratio]{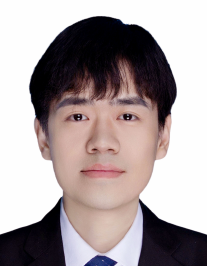}}]{Xu Ma} received the B.S. and M.S. degrees from the School of Computer and Communication Engineering, University of Science and Technology Beijing, Beijing, China, in 2021 and 2024, respectively, where he is currently pursuing the Ph.D. degree. His research interests include satellite networks and resource allocation in next generation wireless communication.
\end{IEEEbiography}
\begin{IEEEbiography}[{\includegraphics[width=1in,height=1.25in,clip,keepaspectratio]{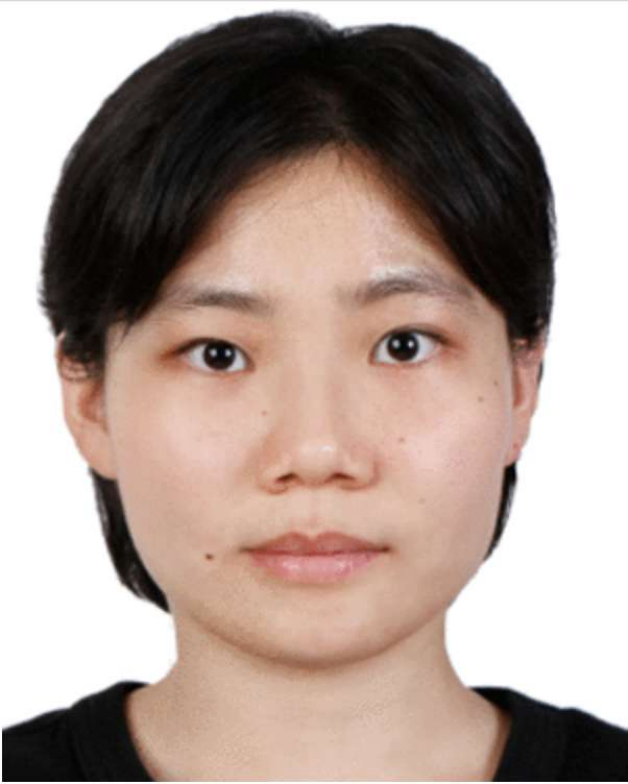}}]{Yuzheng Ren} (Member, IEEE) received the BS and PhD degrees from the Beijing University of Posts and Telecommunications (BUPT), Beijing, China, in 2017 and 2023, respectively. She is currently an associate professor with the School of Computer\&Communication Engineering, University of Science and Technology Beijing. Her research interests include intelligent networks, identity resolution technology, and resource management.
\end{IEEEbiography}

\end{document}